\documentclass[useAMS,usenatbib,a4paper]{mn2e}
\usepackage{color}
\usepackage{graphicx}
\usepackage{breqn}
\usepackage{pgfplots}
\usepackage{epstopdf}
\usepackage{pdflscape}
\usepackage{lscape}
\usepackage{changepage}
\usepackage[maxfloats=40]{morefloats}
\usepackage{rotating}
\newlength\figureheight
\newlength\figurewidth
\usepackage{breqn}
\title[PPTA pulsar timing model parameters]{Timing analysis for 20 millisecond pulsars in the Parkes Pulsar Timing Array}
\author[D. J. Reardon et al.]{D. J. Reardon$^{1,2}$\thanks{E-mail:
Daniel.Reardon@monash.edu}, G. Hobbs$^{2}$, W. Coles$^{3}$, Y. Levin$^{1}$, M. J. Keith$^{4}$, M. Bailes$^{5}$,\newauthor  
N. D. R. Bhat$^{6}$, S. Burke-Spolaor$^{7}$, S. Dai$^{2,8}$, M. Kerr$^{2}$, P. D. Lasky$^{1}$,\newauthor 
R. N. Manchester$^{2}$, S. Os{\l}owski$^{9,10}$, V. Ravi $^{5,11}$, R. M. Shannon$^{2}$,\newauthor  
W. van Straten$^{5}$, L. Toomey$^{2}$, J. Wang$^{12}$, L. Wen$^{13}$, X. P. You$^{14}$, X.-J. Zhu$^{13}$\\
\\
$^{1}$Monash Centre for Astrophysics (MoCA), School of Physics and Astronomy, Monash University, Victoria 3800, Australia\\
$^{2}$Australia Telescope National Facility, CSIRO Astronomy \& Space Science, P.O. Box 76, Epping, NSW 1710, Australia\\
$^{3}$Department of Electrical and Computer Engineering, University of California at San Diego, La Jolla, CA 92093, USA.\\
$^{4}$Jodrell Bank Centre for Astrophysics, University of Manchester, M13 9PL, UK\\
$^{5}$Centre for Astrophysics and Supercomputing, Swinburne University of Technology, P.O. Box 218, Hawthorn, Victoria 3122,\\ Australia.\\
$^{6}$International Centre for Radio Astronomy Research, Curtin University, Bentley, Western Australia 6102, Australia.\\
$^{7}$Jet Propulsion Laboratory, California Institute of Technology, 4800 Oak Grove Drive, Pasadena, CA 91109–8099, USA.\\
$^{8}$School of Physics and State Key Laboratory of Nuclear Physics and Technology, Peking University, Beijing 100871, China\\
$^{9}$Fakult\"{a}t f\"{u}r Physik, Universit\"{a}t Bielefeld, Postfach 100131, 33501 Bielefeld, Germany\\
$^{10}$Max-Planck-Institut f\"{u}r Radioastronomie, Auf dem H\"{u}gel 69, D-53121 Bonn, Germany\\
$^{11}$School of Physics, University of Melbourne, Parkville, Victoria 3010, Australia.\\
$^{12}$Xinjiang Astronomical Observatory, Chinese Academy of Sciences, 150 Science 1-Street, Urumqi, Xinjiang 830011, China.\\
$^{13}$School of Physics, University of Western Australia, Crawley, WA 6009, Australia\\
$^{14}$School of Physical Science and Techology, Southwest University, Chongqing, 400715, China
}

\begin{document}

\date{October 2015}

\pagerange{\pageref{firstpage}--\pageref{lastpage}} \pubyear{2015}

\maketitle

\label{firstpage}

\begin{abstract}
We present timing models for 20 millisecond pulsars in the Parkes Pulsar Timing Array.  The precision of the parameter measurements in these models has been improved over earlier results by using longer data sets and modelling the non-stationary noise. We describe a new noise modelling procedure and demonstrate its effectiveness using simulated data.  Our methodology includes the addition of annual dispersion measure (DM) variations to the timing models of some pulsars. We present the first significant parallax measurements for PSRs J1024$-$0719, J1045$-$4509, J1600$-$3053, J1603$-$7202, and J1730$-$2304, as well as the first significant measurements of some post-Keplerian orbital parameters in six binary pulsars, caused by kinematic effects. Improved Shapiro delay measurements have resulted in much improved pulsar mass measurements, particularly for PSRs J0437$-$4715 and J1909$-$3744 with $M_p=1.44\pm0.07$\,M$_{\odot}$ and $M_p=1.47\pm0.03$\,M$_{\odot}$ respectively. The improved orbital period-derivative measurement for PSR J0437$-$4715 results in a derived distance measurement at the 0.16\% level of precision, $D=156.79 \pm 0.25$\,pc, one of the most fractionally precise distance measurements of any star to date.  

\end{abstract}


\section{Introduction}

The Parkes Pulsar Timing Array (PPTA; Manchester et al. 2013), like its North American  (Demorest et al. 2013) and European (Kramer \& Champion 2013) counterparts, is a program in which an array of millisecond pulsars (MSPs) is observed regularly over many years. The times of arrival (ToAs) of pulses from MSPs are highly predictable using timing models that describe the spin evolution and astrometric properties of the pulsar and any companions, as well as taking into account the motion of Earth and pulse propagation through curved spacetime and the interstellar medium (ISM). The parameters of the timing model are determined, or improved, by a least squares-fit of the model to the ToAs. The differences between the measured and predicted ToAs (the ``timing residuals'') after this fit contain the measurement error, stochastic fluctuations in the apparent pulsar rotation rate (known as timing noise; Shannon \& Cordes 2010), and other unmodelled effects such as those of gravitational-waves (GWs), errors in the assumed time standard, and errors in the solar system ephemeris. Most of these effects are stronger at lower frequencies, i.e., they have a ``red" power spectrum.

The main goal of a pulsar timing array (PTA) is to search for and eventually detect and study nanohertz frequency GWs (e.g. van Haasteren et al. 2011, Demorest et al. 2013, Zhu et al. 2014, and Wang et al. 2015), but there are many secondary objectives such as testing general relativity (GR; e.g. Freire et al. 2012 and Zhu et al. 2015), constraining common models of supermassive black hole and galaxy formation (e.g. Shannon et al. 2015), measuring planetary masses (e.g., Champion et al. 2010), studying the ISM (e.g. You et al. 2007 and Keith et al. 2013; hereafter K13), developing pulsar-based time standards (e.g. Hobbs et al. 2012), and precise measurements of properties of the pulsars themselves (e.g. Verbiest et al. 2008; hereafter V08). The latter includes for example the much improved distance and mass measurements for PSR J0437$-$4715 presented in this paper, which will be important for future limits to changes of Newton's gravitational constant, and the Neutron Star Interior Composition Explorer (NICER) mission that will attempt to measure its radius (Gendreau et al. 2012). The new and improved distance measurements for pulsars presented in this paper are also useful for future galactic electron density models (cf. Cordes \& Lazio 2002).

It has long been known that least-squares fitting for the parameters of the timing model can be biassed and can underestimate the uncertainties on the parameters when the residuals contain significant red noise. To account for this effect, V08 used Monte Carlo simulations for PSR J0437$-$4715 to determine parameter uncertainties. Verbiest et al. (2009; hereafter V09), however, prewhitened the residuals for the three pulsars with the most red noise in their sample by fitting harmonically-related sine/cosine pairs (Hobbs et al. 2006). Coles et al. (2011; hereafter C11) demonstrated that this method can result in biassed parameter measurements and underestimated uncertainties. For all pulsars, V09 then doubled the formal uncertainties obtained from the fit. 

For our work, we use an extension of the ``Cholesky" algorithm developed by C11 and implemented in the timing software package \textsc{tempo2} (Edwards et al. 2006; Hobbs et al. 2006). For the results presented in C11 the red noise was modeled as wide-sense stationary, i.e., having a single power spectrum. However, the algorithm only requires that the red noise be described by a covariance matrix. In our data set the red noise is not stationary because the earlier data contain uncorrected fluctuations in the dispersion measure (DM; the column density of electrons along the line of sight to the pulsar) of the ISM, while the later data do not. Accordingly, we use, and describe below, a modification of the algorithm in C11 that we refer to as the ``split-Cholesky" algorithm, which allows for two different red-noise models in the data set.

In Section 2, we describe the observations and methodology for determining white-noise parameters. In Section 3 we describe the parameters of the timing model. In Section 4 we describe the new split-Cholesky algorithm, modelling of the DM variations, and present parameters describing the red noise and DM noise models. In Section 5, for each pulsar, we present the new timing model parameter values. In Section 6 we present simulations of the split-Cholesky algorithm and compare the method to alternate Bayesian pulsar timing analysis algorithms (e.g. van Haasteren et al. 2011, van Haasteren \& Levin 2013, and Lentati et al. 2014a), and derive a precise pulsar distance for PSR J0437$-$4715.

\section{Observations}
The observations used here were published as the extended first Parkes Pulsar Timing Array (PPTA) data release (DR1E) by Manchester et al. (2013). All observations were taken using the Parkes $64\,\rm{m}$ radio telescope. The data set includes  observations at three observing bands (with approximate centre wavelengths of 10\,cm, 20\,cm, and 50\,cm) from the PPTA project that commenced in 2005 (these observations alone are referred to as data release one; DR1), along with observations prior to 2005 (in the 20\,cm band only) from previous observing programmes.  The earliest data were obtained from a timing programme that commenced during the Parkes 70\,cm survey (Bailes et al. 1994) and were published by Bell et al. (1997) and Toscano et al. (1999). The sample of pulsars was increased by MSPs discovered during the Swinburne intermediate-latitude survey (Edwards et al. 2001) and elsewhere.  Updated timing solutions were published by Hotan et al. (2006) and Ord et al. (2006). Throughout this paper, we refer to the archival 20\,cm observations taken prior to the PPTA as the ``early" data, and the multi-band PPTA observations as ``recent" data.

An intensive observing campaign was used to study PSR J0437$-$4715 in detail. Results were published in van Straten et al. (2001). V08 (for PSR J0437$-$4715)  and V09 (for the other 19 pulsars in the PPTA) combined the earlier data with the initial PPTA data to determine timing ephemerides. Here we use the extra $\sim$3 years of data provided by Manchester et al. (2013) to improve on the results of V08 and V09. Along with the extra data span, our new data set provides significantly improved observing cadence and, for the recent data, the ability to remove the effects of DM variations ($\Delta$DM) more precisely than previously possible. Most of the raw observation files used in this analysis are available from the Parkes pulsar data archive (Hobbs et al. 2011).

Throughout this paper we make use of the \textsc{tempo2} software package to analyse the pulse arrival times (Hobbs et al. 2006). Our analysis method, described below, relies on knowledge of the noise affecting the residuals. Radiometer noise affects all pulsars and is well-modelled by the ToA uncertainty that is obtained from the template-matching procedure carried out when determining the ToA.  However, in almost all cases the observed scatter in the residuals is greater than that expected from radiometer noise alone. This is not unexpected.  Such excess can arise from intrinsic pulse jitter\footnote{For this work, we do not use the jitter parameters introduced by Shannon et al. (2014) because we do not have all of the observation lengths for each ToA in our current dataset.} (e.g., Os{\l}owski et al. 2011, Shannon et al. 2014), calibration errors, instrumental effects, or a poor selection of templates used in the template matching process. \textsc{Tempo2} currently only has two methods for correcting the measured ToA uncertainties: 1) uncertainties for a set of observations can be multiplied by a scaling factor (this is termed an ``\textsc{efac}") or 2) adding a specified amount of extra noise in quadrature with the original uncertainties (termed an ``\textsc{equad}").  If both methods are implemented then the resulting uncertainty is:
\begin{equation}
\sigma_i^\prime = {\rm \textsc{efac}} \times \sqrt{\sigma_i^2 + {\rm \textsc{equad}}^2}
\end{equation}
where $\sigma_i$ is the original uncertainty for the i'th observation.  Determining the \textsc{efac} and/or \textsc{equad} is non-trivial as any low-frequency noise in the residuals must be accounted for and the \textsc{efac} and \textsc{equad} parameters are covariant.  We follow the procedure below for each data set using the \textsc{efacEquad} plugin for \textsc{tempo2} (Wang et al. 2015):
\begin{itemize}
\item Estimate the red noise by fitting a smooth model to the residuals. The default smooth model is a linear interpolation through a set
of samples on 100-day intervals. The smooth model (red noise estimate) is then subtracted from the residuals, leaving only the white
noise.
\item Divide the data into groups based on observing systems that are expected to have the same \textsc{efac} and \textsc{equad} values.  For example, our data set includes data taken using different ``backend" instruments, some of which have identical firmware, bandwidths etc., and are therefore expected to share the same noise properties.
\item For a given group selected from the whitened data set, the reduced-$\chi^2$ value ($\chi^2_r$) is calculated.   If $\chi^2_r < 1$ then ${\rm \textsc{efac}} = \sqrt{\chi^2_r}$ and ${\rm \textsc{equad}} = 0$.  Note that this is the only means to reduce a ToA uncertainty as an \textsc{equad} will always increase the uncertainty.
\item  If $\chi^2_r > 1$ then the normalised residuals ($r_i/\sigma_i$) are determined for a particular grid of \textsc{efac} and \textsc{equad} values.  
\item For each grid position, we determine the probability that the normalised residuals are drawn from a Gaussian distribution (using a Kolmogorov-Smirnov test) to determine optimal \textsc{efac} and/or \textsc{equad}.
\item By default, we do not include the \textsc{efac}s and \textsc{equad}s for a group that consists of less than 10 ToAs. They are added if it is necessary to produce normally-distributed residuals.
\end{itemize}

\section{The timing model}

The timing model describes the spin, astrometric, and orbital properties of a pulsar, the ISM along the line of sight, and requires the use of a terrestrial time standard, solar system ephemeris, and solar wind model. For this work we use the DE421 Jet Propulsion Laboratory solar system ephemeris and as a time reference use TT(BIPM2013). We use the default model within \textsc{tempo2} to account for dispersion measure variations caused by the solar wind (wind density at 1 A.U.: 4~cm$^{-3}$; Edwards et al. 2006).  The pulsars have been observed over many years with various different ``backend" instruments.  These instruments have different time offsets which we also measure as part of the usual timing fit (with the split-Cholesky method).

The thirteen pulsars in our sample that are in binary systems each have a white dwarf companion.  For such systems, we do not expect any time dependencies for the orbital parameters caused by mass loss or spin-orbit coupling.  Because of the relatively low mass of the companion stars and relatively long orbital periods, relativistic effects are small in such systems. However, V08 do report a detection of the advance of the longitude of periastron, $\dot{\omega}$, for PSR J0437$-$4715 that is consistent with that predicted from GR, where the component masses were derived from the Shapiro delay measurement and mass function (Thorsett \& Chakrabarty 1999). van Straten (2013) also measures $\dot{\omega}$ for PSR J1022$-$1001 that is consistent with GR.

For binary pulsars in orbits with small eccentricities, the longitude and epoch of periastron are not well defined and are highly correlated. The ELL1 model (Lange et al. 2001) is used to describe such systems, since it uses a small-eccentricity approximation to avoid this high correlation. For PSRs J1022+1001, J1600$-$3053, and J1643$-$1224 this approximation is not valid and so we instead use the DD (Damour \& Deruelle 1986) model to describe the binary orbit.

Observed changes in the orbital parameters can be caused by kinematic effects.  Changes in the apparent viewing geometry of the orbit caused by proper motion can lead to an apparent time derivative of the projected semi-major axis of the orbit, $\dot{x}$, and/or $\dot{\omega}$ (Kopeikin 1996). For some pulsars, this kinematic $\dot{x}$ or $\dot{\omega}$ may be detected individually. However, if both are well determined we instead parametrise the effect with $\Omega$ and $i$, which are the longitude of the ascending node and inclination angle of the orbit respectively (van Straten \& Bailes 2003). This parametrisation of the orbit describes the annual orbital parallax and is implemented through the T2 model (Edward et al. 2006), which we use for PSRs J0437$-$4715, J1713$+$0747, and J1909$-$3744. For PSRs J1022+1001, J1600$-$3053, J1603$-$7202, J1643$-$1224, and J2145$-$0750 we use the $\dot{x}$ measurement to place an upper limit on the inclination angle of the orbit (using $\tan{i}\leq x\mu/\dot{x}$, where $\mu$ is the total proper motion), as was done by Sandhu et al. (1997).

A pulsar can have an apparent spin frequency derivative ($\dot{\nu}$) or orbital period-derivative ($\dot{P}_b$) as a result of the Shklovskii effect (Shklovskii, 1970); an apparent radial acceleration of the system caused by proper motion perpendicular to the line of sight.  As we do not include this explicitly in the timing model, we expect non-zero values for these parameters.  The $\dot{\nu}$ value caused by the Shklovskii effect is simply absorbed into the intrinsic spin-down rate for the pulsar.  However, if the expected intrinsic $\dot{P}_b = 0$ then as shown by Bell \& Bailes (1996) the observed value can be used to determine the distance to the pulsar:
\begin{equation}
D=\frac{c}{\mu^{2}}\frac{\dot{P}_b^{\rm obs}}{P_b}
\end{equation}
where $D$ is the pulsar's distance, $c$ the vacuum speed of light, and $\mu$ the total proper motion. This equation neglects the differential acceleration of the pulsar and the Earth in the gravitational potential of the Galaxy. The pulsar's distance can therefore be determined from $\dot{P}_b^{\rm obs}$ or from the annual parallax when these effects are taken into account.  We choose to decouple these parameters (i.e., fit separately for the two distances) in order to determine whether the distance measurements are consistent. The pulsar timing model requires a distance estimate when determining the orbital parallax and the annual orbital parallax (Kopeikin 1995). In all cases here, we use the default parallax distance. In Section 6.2 we discuss the significance of this $\dot{P}_b$ distance measurement for PSR J0437$-$4715.

\subsection{Choosing parameters to include in the model}
For all pulsars we start with the timing models presented by V09 (and, for PSR J0437$-$4715, the model from V08). For solitary pulsars, model parameters include the spin ($\nu$, $\dot{\nu}$), astrometric (position, proper motion, and parallax), and ISM parameters (Section 4.1). In all cases, we fit for the spin, position (right ascension $\alpha$, and declination $\delta$), and DM variation parameters (from Manchester et al. 2013). These parameters are all measured at a reference epoch of MJD 54500. In a few cases (in particular where the pulsar is either close to or almost perpendicular to the ecliptic plane respectively) we cannot obtain a significant measurement of the proper motion in declination ($\mu_\delta$) or the parallax ($\pi$).

To determine whether parameters beyond this base model are required by the data, we make use of the Akaike information criterion (AIC; Akaike 1973), which states that a model is a better fit to the data if 
\begin{equation}
\Delta \chi^2 > 2k
\end{equation}
where $\Delta \chi^2$ is the difference of the $\chi^2$ value before and after a fit that includes $k$ new parameters. To determine which parameters to fit for using the AIC, we use the following procedure with solitary pulsars:
\begin{itemize}
\item{Remove non-essential parameters (proper motion in right ascension $\mu_\alpha$ and declination $\mu_\delta$, and parallax $\pi$) from the timing model (if present) and note the $\chi^{2}$ value of residuals.}
\item{Fit for each parameter separately and note the new $\chi^{2}$ value in each case.}
\item{Include the parameter that results in the lowest $\chi^{2}$ value permanently into the timing model if this parameter also satisfies the AIC.}
\item{Repeat with remaining parameters until either all parameters are included in the timing model, or all remaining parameters fail to improve the timing solution, as determined by the AIC.}
\end{itemize}

This procedure is also applied to pulsars in binary systems. All non-essential binary parameters (post-Keplerian) are initially removed and the AIC is used to find which should be included. If a Shapiro delay (Shapiro 1964) is detectable for the binary system, we parametrise this with the companion mass $M_{\rm c}$, and sine of inclination angle of the orbit, $\sin i$, except in the cases of pulsars described with the T2 binary model, for which we link $\sin i$ with the measured inclination angle from the Kopeikin terms (Kopeikin 1996).

\section{Parameter measurement in the presence of non-stationary red noise}

In the original implementation of the C11 algorithm, the red noise was characterized by a power spectrum, from which a covariance function was estimated and finally a covariance matrix was constructed. This was satisfactory for analysis of the DR1 data set. However, DR1E data contain uncorrected DM variations in the 20\,cm residuals prior to multi-band observations and thus have additional red noise. We cannot produce a single power-law model for the entire data set because of this extra red noise which may dominate the total red noise in a subset of the data. Instead we produce two separate models to describe the two sources of red noise in the residuals. One model describes the frequency-independent noise present throughout the dataset using a power-law, while the other describes the additional DM noise present only in the early data (Section 4.1). We have updated the implementation of the algorithm to synthesize a covariance matrix for DR1E observations using these two red-noise models, which we refer to as the split-Cholesky algorithm.

The method requires two red noise covariance matrices, one for the frequency-independent timing noise and another for the DM noise. Both of these are estimated from the DR1 data alone because for this data set we can estimate and remove the DM noise (described in Section 4.1), allowing us to model the timing noise (Section 4.2) independently with an analytical model,

\begin{equation}
P(f) = \frac{P_0}{\left[1+\left(\frac{f}{f_c}\right)^2 \right]^{\alpha/2}}
\label{eqn:RedNoise}
\end{equation}
where $P_0$ is the amplitude of the power in ${\rm yr}^3$ at a corner frequency $f_c$, and $\alpha$ is the spectral index. A stationary covariance matrix is then computed from this spectral model. 

To build the DM noise model that applies to only the early DM-uncorrected observations, we first estimate DM variations, $\Delta$DM(t) from the DM-corrected DR1 data using a process described in Section 4.1. We can then choose to create an analytical model of the power spectrum of $\Delta$DM(t) from which we compute a covariance function $C_f(\tau)$ (Keith et al. 2013), or we can model the covariance function of $\Delta$DM(t) directly (Section 4.1). To create the final covariance matrix for the entire data set we must account for the fact that we do not know the mean $\Delta$DM for the DR1 DM-corrected data. It is adjusted to match the end of the earlier uncorrected data with no discontinuity. We then compute the non-stationary $\Delta$DM(t) covariance matrix $C_m(t_i, t_j)$ as follows:
$$C_m(t_i, t_j) = C_f(|t_i - t_j|) $$    for $t_i, t_j < T_c$
$$C_m(t_i, t_j) = C_m(T_c, T_c) = C_f(0)   $$   for $t_i, t_j > T_c$, and
$$C_m(t_i, t_j) = C_f(|t_i - T_c|) $$    for $t_i < T_c$, $t_j > T_c$,
where $T_c$ is MJD 53430, the time in the dataset beyond which the data is DM corrected. Finally we sum the stationary covariance matrix for the red timing noise (C11), the non-stationary covariance matrix for the $\Delta$DM(t), and the diagonal matrix of the variances of the white noise at each sample and apply the Cholesky algorithm as originally formulated. In Section 6.1 we demonstrate the effectiveness of this algorithm through the use of simulated data.

\subsection{Modelling the Dispersion Measure variations}

With the advent of the PPTA in 2005, regular observations occurred in multiple observing bands (10cm, 20cm, and 50cm). For these data we are, in principle, able to obtain a measurement of the DM because the group delay is $\propto \lambda^2$. However, as the pulse profiles of MSPs in our sample evolve significantly with frequency (Dai et al. 2015), the effect of an absolute DM on the residuals is coupled with the frequency evolution of the pulse profile and is therefore difficult to determine. We are however able to measure changes in DM, which we refer to as $\Delta$DM(t), using interband measurements with the required accuracy ($\approx 1:10^5$). Residual errors in these corrections are an important source of noise in the PPTA (and all other PTAs). Such time series (obtained from the same observations we analyse here but with different EFACs and EQUADs) were analysed by K13.

We assume that the $\Delta$DM(t) are caused by the movement of the line of sight from the Earth to the pulsar through spatial variations in the ISM. If the velocity of the line of sight were constant, $\Delta$DM(t) would simply represent a cut through the ISM in the direction of the velocity. If the fluctuations are due to homogeneous Kolmogorov turbulence then the power spectrum of $\Delta$DM(t) would be (K13)
\begin{equation}
P_{\rm DM}(f) \simeq 3.539D(\tau)\tau^{-5/3}f^{-8/3}
\end{equation}
where $D(\tau)$ is the structure function at time lag $\tau$. Here $D(\tau)$ is measured in s$^{2}$, $\tau$ in s, $f$ in yr$^{-1}$, and $P_{\rm DM}(f)$ in yr$^{3}$.

Many of the pulsars in K13 showed a clear linear trend in $\Delta$DM(t), possibly indicating a constant spatial gradient over the observing span.  In such cases the Earth's orbital motion causes an annual sinusoid in $\Delta$DM(t) and this was also observed by K13. Although this gradient may be part of a stochastic process, for the purpose of analysing our observations it can be considered deterministic and included in the timing model. Accordingly, if it is statistically significant (determined using an AIC test), we fit and remove a linear gradient (dDM/dt) and an annual sinusoid that has been added to the parameters of the timing model in \textsc{tempo2} with the equation
\begin{equation}
{\rm{DM_{yr}}}=A\sin\left(2\pi {\rm yr}^{-1}\left(t - T_0\right)\right) + B\cos\left(2\pi {\rm yr}^{-1}\left(t - T_0\right)\right)
\end{equation}
where $A$ and $B$ are the parameters in the timing model that describe the amplitude and phase of the DM annual variations and $T_0$ is the reference epoch for the DM measurements. If $\Delta$DM(t) has a linear trend and/or annual variation, which we include in the timing model using dDM/dt, and/or $A$ and $B$, then we need to measure and model the covariance, $\textrm{Cov}(\tau)$, of the residual $\Delta$DM(t). 

We measure $\Delta$DM(t) with a 5\,yr$^{-1}$ cadence for each pulsar using the DR1 multi-band data, and convert each measurement to a time delay in the 20\,cm band using $t_{\rm DM}(\nu) = {\rm DM}/(K\nu^2)$, where $\nu$ is the observing frequency (1400\,MHz in this case) and $K=2.410\times10^{-4}$\,MHz$^{-2}$cm$^{-3}$pc\,s$^{-1}$ (You et al. 2007). We then model the covariance functions of the detrended (if required) $\Delta$DM(t) with a function of the form
\begin{equation}
\textrm{Cov}\left(\textrm{DM}(t)\right) = a\exp{\left(-{\left(\frac{\tau}{b}\right)}^{\alpha}\right)}.
\end{equation}
The covariance (Cov) is a function of lag $\tau$ (in days), $a$ is the amplitude of the red noise (in s$^2$), $b$ is the characteristic timescale (in days), and the exponent was chosen to be $\alpha=2$ so that the covariance function will have a positive definite Fourier transform, which is the power spectrum. 

The annual variation, linear trend, and covariance function parameters that we have used to construct a DM model for each pulsar are given in Table \ref{tab:DMparam}. In some cases, no dDM/dt is apparent in the data, but the $\Delta$DM(t) noise is nevertheless well modelled by the covariance function since it is small. For all pulsars with the exception of PSR J0437$-$4715 (which is very well modelled by a Kolmogorov power law), we find that these covariance function models successfully whiten the residuals and we therefore include them in our combined red-noise models. For PSR J0437$-$4715 we use the Kolmogorov power law presented in K13 to model the DM noise present in the early data.

For PSR J1603$-$7202 there is an extreme scattering event (ESE; Fiedler et al. 1987), lasting $\sim$250 days, which was reported by K13 and is described in detail in Coles et al. (2015). The ESE dominates the shape of the non-DM-corrected 20\,cm residuals in that region. We examined the early 20\,cm data searching for ESEs comparable with the one reported by K13 and found none. We obtained the covariance model for $\Delta$DM(t) by linearly interpolating across the ESE before computing the covariance function. 

Maitia \& Lestrade (2003) reported on a 3-year-long ESE (centred on the year 1998) detected in the direction of PSR J1643$-$1224 by studying flux variability of the pulsar using observations undertaken at the Nancay observatory. Unfortunately we have poor data during this time and consequently do not find evidence for such an event. For PSR J1713$+$0747 we see a peculiar ``drop out" in the $\Delta$DM(t) that is probably related to an ESE-like structure. We therefore computed the covariance of $\Delta$DM(t) for PSR J1713$+$0747 by interpolating across this drop out.

The deterministic terms in the model for each pulsar include an absolute DM measurement (which we hold constant because it is covariant with pulse profile evolution; Dai et al. 2015), and if required, the additional terms dDM/dt, $A$ and $B$, or $\Delta$DM(t) measurements taken at a spacing of $\Delta t_{\rm{DM}}$ days to remove any residual red noise. The sampling interval $\Delta t_{\rm{DM}}$ is ideally the widest spacing required to remove the red noise since the measurements add white noise to the residuals. We start with the  $\Delta t_{\rm{DM}}$ published with the dataset in Manchester et al. (2013). We sample more frequently if there is residual red noise because the published $\Delta t_{\rm{DM}}$ were selected to minimise the rms residual in the best band for each pulsar, and not to absorb the most red noise. However adding the deterministic DM parameters often allows us to reduce the $\Delta$DM(t) measurement cadence relative to that published in Manchester et al. (2013).

\begin{table*}
\begin{center}
\caption{Parameters describing the DM model used for each pulsar. DM noise in the earliest residuals is described by the DM covariance function parameters ($a$ and $b$) in Equation 7, which are calculated after the removal (if necessary) of a linear trend, dDM/dt, and annual variations described by the sine (A) and cosine (B) amplitudes. $\Delta t_{\rm{DM}}$ is the separation of $\Delta \rm{DM}$ measurements in the multi-frequency section of the dataset for each pulsar.}
\begin{tabular}{llllllll}
\hline
\hline
& \multicolumn{5}{c}{Timing model parameters} & \multicolumn{2}{c}{DM covariance}\\
 & DM & dDM/dt & $A$ & $B$ &$\Delta t_{\rm{DM}}$ &$a$  & $b$ \\
  Pulsar Name & (cm$^{-3}$pc) &(cm$^{-3}$pc yr$^{-1}$) & (10$^{-4}$\,cm$^{-3}$pc)&(10$^{-4}$\,cm$^{-3}$pc)&(days)&(s$^2$)&(days)\\
\hline
J0437$-$4715$^*$ & 2.64498 & -- & -- & -- & 60 & -- &-- \\
J0613$-$0200 & 38.7756 & --& $-1.0\pm0.2$& $-1.0\pm0.2$ & 365.25 & 5.5$\times 10^{-14}$ & 317\\
J0711$-$6830 & 18.4099 & $(9\pm7)\times 10^{-5}$& --& -- & 200 & 5.8$\times 10^{-13}$ & 331 \\
J1022$+$1001 & 10.2531 & -- & --& -- & 200& 1.2$\times10^{-13}$ & 153\\
J1024$-$0719 & 6.48803 & $(2.2\pm0.6)\times 10^{-4}$& -- & --& -- & -- & -- \\
J1045$-$4509 & 58.1438 & $(-3.66\pm0.13)\times 10^{-3}$& $-8.1\pm2.3$& $-0.9\pm4$ & 182.62 & 1.7$\times 10^{-11}$ & 179\\
J1600$-$3053 & 52.3249 & $(-6.3\pm0.3)\times 10^{-4}$ & --& -- & 125& 5.2$\times 10^{-13}$ & 146\\
J1603$-$7202 & 38.0489 & -- & --& -- & 100& 6.3$\times 10^{-13}$ & 64 \\
J1643$-$1224 & 62.4143 & $(-1.23\pm0.005)\times 10^{-3}$& $-2.9\pm0.7$& $-5.9\pm0.7$ & 365.25 & 1.5$\times 10^{-12}$ & 113\\
J1713$+$0747 & 15.9903 & -- & --& -- & 365.25& 2.6$\times 10^{-14}$ & 171 \\
J1730$-$2304 & 9.61634 & $(5.6\pm0.5)\times 10^{-4}$ & -- & --& --& -- & --\\
J1732$-$5049 & 56.8365 & $(8.8\pm1.2)\times 10^{-4}$& -- & --& --& -- & -- \\
J1744$-$1134 & 3.13695 & $(-1.32\pm0.18)\times 10^{-4}$ & --& -- & --& -- & -- \\
J1824$-$2452A & 119.892 & $(1.15\pm0.08)\times 10^{-3}$& -- & --& 82.5 & -- & -- \\
J1857$+$0943 &  13.2984 & $(2.8\pm0.5)\times 10^{-4}$& -- & --& -- & -- & -- \\
J1909$-$3744 & 10.3932 & $(-2.97\pm0.06)\times 10^{-4}$& -- & --& 105 & -- & --\\
J1939$+$2134 & 71.0227 & $(-5.9\pm0.3)\times 10^{-4}$ & $2.4\pm1.1$& $1.6\pm1.2$ & 50& 3.2$\times 10^{-13}$ & 112 \\
J2124$-$3358 & 4.60096 & --& --& -- & -- & -- & --\\
J2129$-$5721 & 31.8509 & $(-1.6\pm0.4)\times 10^{-4}$& --& -- & -- & -- & -- \\
J2145$-$0750 & 8.99761 & $(1.2\pm0.3)\times 10^{-4}$& -- & --& -- & -- & -- \\
\hline
\end{tabular}
\newline
$^*$ Kolmogorov model from K13 is used to model the DM noise instead of a covariance function. \\
\label{tab:DMparam}
\end{center}
\end{table*}

\subsection{Modelling the red timing noise}
The analytical model (Equation \ref{eqn:RedNoise}) for the frequency-independent red noise (e.g. timing noise) must be estimated with data that have the frequency-dependent red noise contributions from DM variations already removed. We therefore use the DM-corrected DR1 data and fit the analytical model (Equation \ref{eqn:RedNoise}) to the power spectrum. In most cases we find that $f_c = 1/T_{1}$ fits the data reasonably well, where $T_{1}$ is the length of the DR1 data. For these cases we assume that the model will fit the entire DR1E with $f_{c} = 1/T_{E}$, where $T_{E}$ is the length of the DR1E dataset, and we scale $P_0$ to this new $f_{c}$ accordingly.

In many cases, extrapolating the timing noise model by adjusting $f_c$ will underestimate the noise at $f \approx 1/T_{E}$ because fitting for the spin frequency and its derivative removes much of the low-frequency power at $f \approx 1/T_{1}$. For pulsars where the timing noise dominates the DM noise, we create a single spectral model for the entire DR1E span directly. If this model ultimately whitens the residuals adequately we do not need to estimate the covariance matrix of the DM variations. If however the timing noise model does not extrapolate well and the DM noise is too significant to ignore, we construct a red-noise model using the DR1E data, but reduce the amplitude, $P_0$, to account for the known DM noise in the data. The amplitude need only be reduced such that the residuals are sufficiently whitened by the red-noise model.

For PSRs J0613$-$0200, J1600$-$3053, and J2145$-$0750 we find that the red-noise model from the DR1 data alone does not extrapolate well over the entire dataset since the power spectrum of the DR1E data is observed to turn over at $f_c > 1/T_{E}$. This was not obvious when analysing the DR1 data alone. We observe a similar turn over for PSR J1909$-$3744, however the dataset used for this pulsar includes multi-band observations across the entire span and so is largely DM-corrected (see Section 5.16).  For this pulsar, we therefore model the power spectrum of the entire data set and use this with the original Cholesky algorithm.

The parameters describing the red timing noise models for each pulsar are given in Table~\ref{tab:RedNoiseparam}. The DM models that are used with these red-noise models are described in Section 4.1, with parameters in Table \ref{tab:DMparam}.

\begin{table}
\begin{center}
\caption{Parameters for the red-noise model for each pulsar. The parameters are used to describe the frequency-independent noise in the data with a power law (Equation 4), where $\alpha$ is the spectral index and $P_0$ is the power at corner frequency, $f_c$. }
\begin{tabular}{llll}
\hline\hline
 Pulsar Name & $\alpha$ & $P_0$ (yr$^3$)& $f_c$ (yr$^{-1}$)\\
\hline
J0437$-$4715 & 3 & $1.14\times 10^{-27}$ & $0.067$ \\
J0613$-$0200 & 5 & $5.5\times 10^{-28}$ & $0.40$  \\
J0711$-$6830 & -- & -- & --\\
J1022$+$1001 & -- & -- & -- \\
J1024$-$0719 & 6 & $1.8\times 10^{-23}$ & $0.066$\\
J1045$-$4509 & 3 & $2.0\times 10^{-24}$ & $0.059$ \\
J1600$-$3053 & 2.5 & $3.0\times 10^{-28}$ & $0.40$ \\
J1603$-$7202  & 2.5 & $1.2\times 10^{-25}$ & $0.065$\\
J1643$-$1224 & 4 & $1.5\times 10^{-25}$ & $0.15$ \\
J1713$+$0747 & 2 & $3.0\times 10^{-27}$ & $0.059$ \\
J1730$-$2304 & -- & -- & -- \\
J1732$-$5049 & 2 & $3.0\times 10^{-27}$ & $0.25$\\
J1744$-$1134 & -- & -- & --\\
J1824$-$2452A & 3.5 & $3.0\times 10^{-25}$ & $0.17$\\
J1857$+$0943 & -- & -- & --\\
J1909$-$3744 & 2 & $1.2\times 10^{-29}$ & $0.50$ \\
J1939$+$2134 & 4.5 & $1.5\times 10^{-24}$ & $0.064$ \\
J2124$-$3358 & 3.5 & $2.0\times 10^{-25}$ & $0.06$ \\
J2129$-$5721 & 1 & $1.0\times 10^{-27}$ & $0.065$\\
J2145$-$0750 & 4 & $3.0\times 10^{-26}$ & $0.3$\\
\hline
\end{tabular}
\label{tab:RedNoiseparam}
\end{center}
\end{table}

\subsection{How do we know when our models are optimal?}
The split-Cholesky algorithm produces the best linear unbiased estimators if, and only if, the residuals after whitening are white and normally distributed. If this is not the case, one or more of the red-noise model, \textsc{efac}s, or \textsc{equad}s are incorrect. We utilise two tests to check that the final residuals are what we require to have confidence in the parameter measurements and uncertainties.

\begin{figure*}
\centering
\includegraphics[trim=100 50 100 50,clip,width=\textwidth]{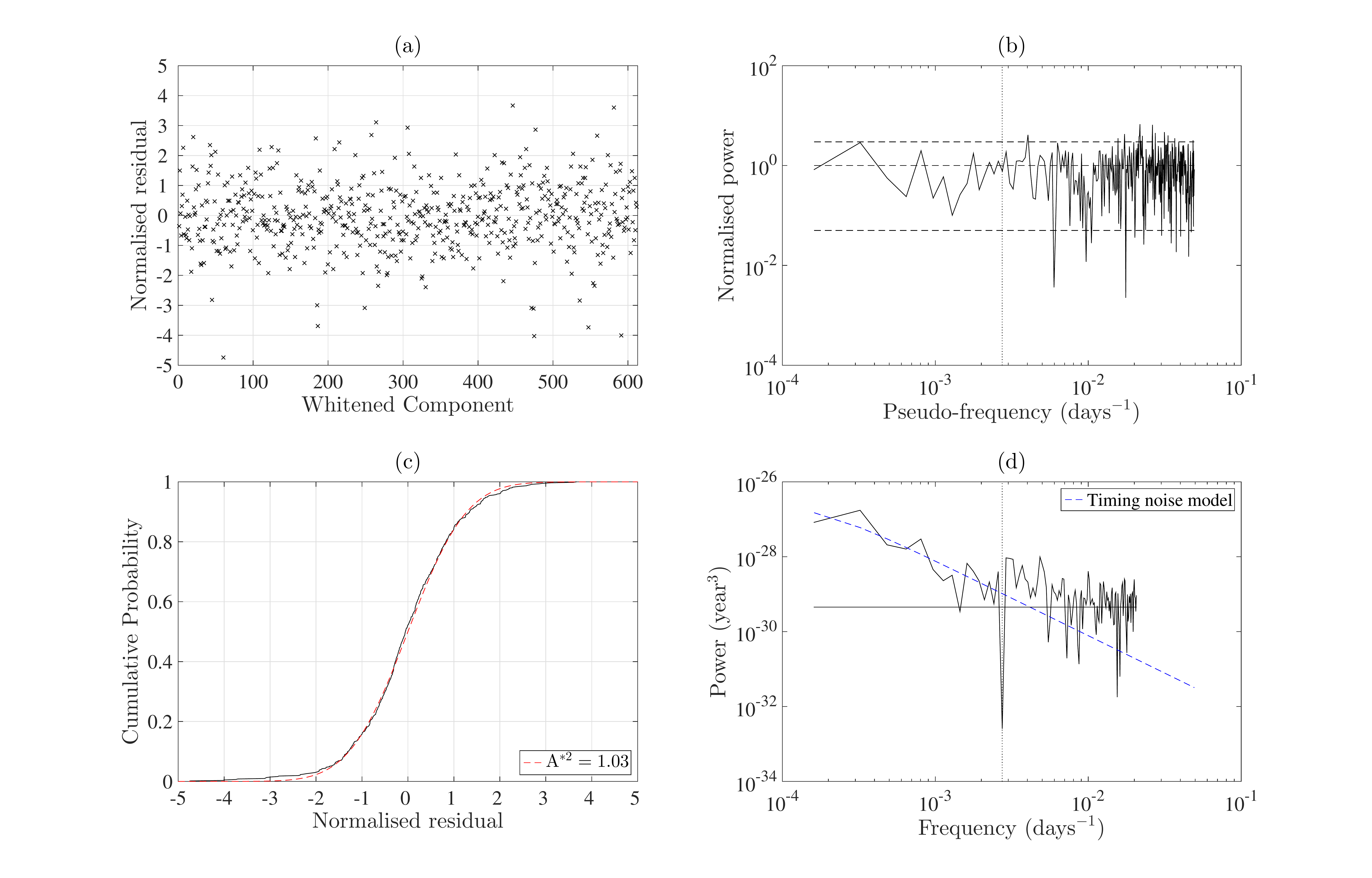}
\caption{(a) Whitened and normalised post-fit residuals for PSR J1713+0747. (b) Power spectra of whitened and normalised post-fit residuals. Dotted line marks $f=1$\,yr$^{-1}$ and dashed lines indicate expected mean and variance for the spectrum. The pseudo-frequency is determined by converting the whitened components to a pseudo-time-series using the ToAs of the unwhitened residuals. (c) Cumulative distribution of whitened and normalised post-fit residuals (solid line) with expected distribution based on normal distribution with zero mean and unit variance (dashed line). Modified AD statistic for this distribution with the expected distribution is labelled. (d) Power spectra of post-fit residuals. Dotted line marks $f=1$\,yr$^{-1}$ and solid flat line is an estimate of the white-noise level. Dashed line is the power law model of the frequency-independent timing noise. }
\label{fig:J1713Analysis}
\end{figure*}

The first of these is an Anderson-Darling (AD) test for normality (Anderson \& Darling 1954). We apply this to our whitened, normalised residuals to determine whether they are consistent with a normal distribution with $\mu=0$ and $\sigma=1$. The result is the modified AD statistic, $A^{*2}$. This is used to test the hypothesis that the residuals obey the described normal distribution. The hypothesis is rejected if $A^{*2} > 2.492$ with 5\% significance (Stephens 1974), since the expected distribution function for the normalised residuals is known. For example, panel (a) of Figure \ref{fig:J1713Analysis} shows the post-fit whitened and normalised residuals for PSR J1713+0747 and panel (c) shows the cumulative distribution of these residuals. The modified AD statistic, $A^{*2} = 1.03$, indicates that the normalised residuals are consistent with a standard normal distribution.

We then test for ``whiteness" by inspecting the power spectrum of the whitened, normalised residuals (see also C11). Panel (b) of Figure \ref{fig:J1713Analysis} shows the Lomb-Scargle periodogram of these residuals for PSR J1713+0747. We compute this power spectrum by first converting the whitened components in panel (a) to a time series using the ToAs of the unwhitened residuals. The frequency axis of the resulting power spectrum is not well-defined, and we term it the pseudo-frequency. However the whitening process provides a diagonally-dominant whitening matrix such that low frequencies in the unwhitened residuals translate to low frequencies in the pseudo-time-series of the whitened components. The pseudo-frequency power spectrum can therefore be a useful test of the red-noise model used in the whitening process, since it must be flat. The power spectrum of the unwhitened residuals is shown in panel (d) of Figure \ref{fig:J1713Analysis} with the timing noise model used for this pulsar. If the test fails for any pulsar, we update the red-noise model and re-fit the timing model until the residuals are successfully whitened as required.

\section{Results}
\begin{figure*}
\centering
\includegraphics[trim=80 20 25 10,clip,width=35pc]{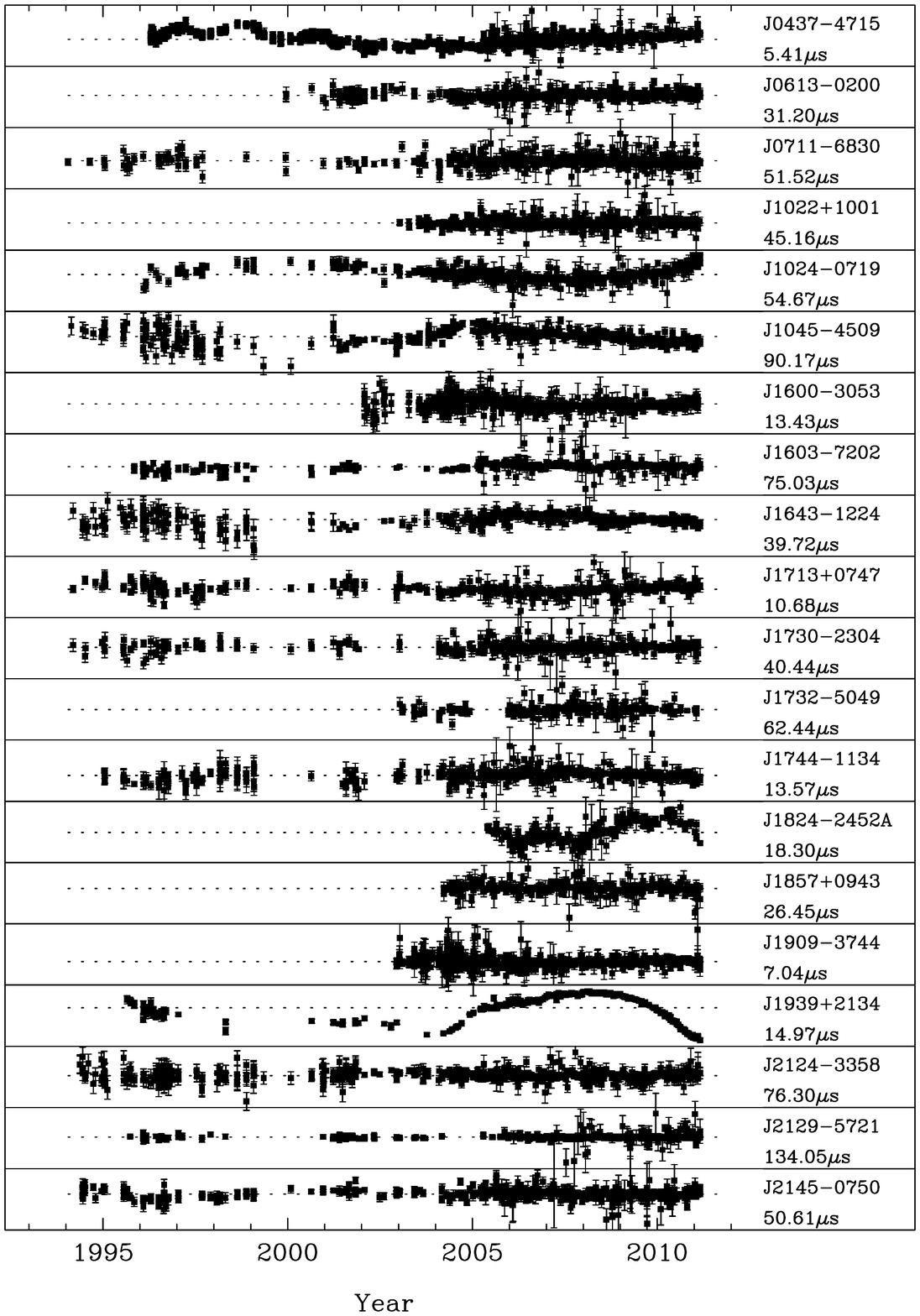}
\caption{Final post-fit residuals for each of the pulsars in our sample. The vertical range of each subplot is given below the pulsar name.}
\label{fig:residuals.pdf}
\end{figure*}

In this section, we present our final timing solutions for each pulsar, with the post-fit residuals given in Figure~\ref{fig:residuals.pdf}. We compare our measured parameters with those in the literature, including VLBI measurements and distances derived from the pulsar DMs using the Taylor \& Cordes (1993; hereafter TC93) and NE2001 (Cordes \& Lazio 2002) galactic free electron distribution models (giving distances accurate to approximately 25\% and 20\% respectively). 

Much of the comparison in this section will be with V08 and V09, which used a subset of our dataset and is therefore not independent. However, V08 and V09 used a different solar system ephemeris, DE405, and time standard, TT(TAI), to our analysis, which results in apparent changes to some parameters. Use of the DE405 ephemeris in particular induces significantly different position parameters compared to the newer DE421 ephemeris, while the use of TT(TAI) changes the apparent spin frequency and its derivative. For each pulsar we find that inconsistencies in the position parameters are explained by the use of this different ephemeris.

Although Keith et al. (2013) measured DM variations for all PPTA pulsars and determined an optimal sampling time, Manchester et al. (2013) did not publish any $\Delta$DM measurements for seven pulsars. This is because $\Delta$DM values were only published by Manchester et al. (2013) if their inclusion improved the rms residuals. For our work, we must account for all red noise in the residuals since this is required for the split-Cholesky algorithm (this process is described in Section 4). The testing of red- and white-noise models for each pulsar is described in Section 4.3, where we present the analysis of PSR J1713+0747 as an example. We do not present the analysis of the other pulsars, with the exception of PSR J0437$-$4715, which shows some variation that we discuss below in Section 5.1. 

For each pulsar we also compare the various distance measurements available in the literature to check that our new distance measurements from $\pi$ or $\dot{P_b}$ are consistent with these. For each parallax measurement, we calculate the Lutz-Kelker (Lutz \& Kelker 1973) corrected parallax value and the corresponding corrected distance using the method of Verbiest et al. (2012) and with the mean flux of each pulsar given in Manchester et al. (2013). The parallax measurements for PSRs J1024$-$0719, J1045$-$4509, J1603$-$7202, and J1857+0943 have high fractional uncertainties (parallax measured with less than 3-$\sigma$ confidence) and therefore the bias corrected values for these closely resemble the prior distributions used. The measurements however are important for placing an upper-bound on the parallax value, or similarly, a lower-bound on the distance measurement. A table of these values, and the distances from $\dot{P_b}$ and DM are given in Table \ref{tab:Distances}.

For each pulsar, we derive the characteristic age, surface magnetic field strength, and energy-loss rate from the spin-down. Where available, we use the pulsar mass function (Thorsett \& Chakrabarty 1999) and a precise Shapiro delay measurement to calculate a pulsar mass, and use the kinematic $\dot{x}$ measurements to place a limit on the inclination angle of the orbit. These derived parameters are given in the Appendix with the tables of parameters.

The red-noise models, ToA files, and parameter files for each pulsar are available from the The Commonwealth Scientific and Industrial Research Organisation (CSIRO) Data Access Portal\footnote{http://dx.doi.org/10.4225/08/561EFD72D0409}.

\begin{table*}
\begin{center}
\caption{Parallax values and distance measurements for pulsars in our sample. Parallax and parallax-derived distance values are corrected for the Lutz-Kelker bias using the method of Verbiest et al. (2012). TC93 and NE2001 DM distances have approximate uncertainties of 25\% and 20\% respectively. }
\begin{tabular}{lllllll}
\hline
\hline
& \multicolumn{2}{c}{Measured values} & \multicolumn{2}{c}{L-K bias corrected values} & \multicolumn{2}{c}{DM distances (kpc)}\\
Pulsar Name & Parallax $\pi$ (mas)& $\dot{P_b}$ distance (kpc) & Parallax $\pi$ (mas) & $\pi$ distance (kpc) & TC93 & NE2001 \\
\hline
J0437$-$4715 & $6.37\pm0.09$ & $0.15679\pm0.00025$ & $6.37\pm0.09$ & $0.156.9\pm0.0022$  & 0.14 & 0.189 \\
J0613$-$0200 & $0.86\pm0.13$ & -- & $0.81\pm0.13$ & $1.09^{+0.18}_{-0.14}$  & 2.19 & 1.700 \\
J0711$-$6830 & -- & -- & -- & -- & 1.04 & 0.854 \\
J1022$+$1001 & $1.1\pm0.3$& $1.2\pm0.5$ & $1.0\pm0.3$ & $0.74^{+0.19}_{-0.13}$ & 0.60 & 0.443 \\
J1024$-$0719 & $0.5\pm0.3$& -- &  $0.5^{+0.21}_{-0.16}$ & $1.1^{+0.4}_{-0.3}$  & 0.35 & 0.381 \\
J1045$-$4509 & $2.2\pm1.1$ & --& $0.29^{+0.5}_{-0.15}$ & $0.34^{+0.2}_{-0.10}$  & 3.24 & 1.945 \\
J1600$-$3053 & $0.48\pm0.11$ & --& $0.43\pm0.11$ & $1.8^{+0.5}_{-0.3}$ & 2.67 & 1.581 \\
J1603$-$7202 & $1.1\pm0.8$ & $3.9\pm1.8$& $0.25^{+0.4}_{-0.12}$ & $0.53^{+0.4}_{-0.16}$  & 1.64 & 1.159 \\
J1643$-$1224 & $1.27\pm0.19$ & --& $1.18\pm0.19$ & $0.74^{+0.12}_{-0.10}$  & $>$4.86 & 2.320 \\
J1713$+$0747 & $0.86\pm0.09$ & $3.1\pm1.2$& $0.84\pm0.09$ & $1.12^{+0.12}_{-0.11}$  & 0.89 & 0.889 \\
J1730$-$2304 & $1.5\pm0.3$ & --& $1.2\pm0.4$ & $0.62^{+0.15}_{-0.10}$  & 0.51 & 0.529 \\
J1732$-$5049 & -- & -- & -- & -- & 1.81 & 1.392 \\
J1744$-$1134 & $2.53\pm0.07$ & --& $2.52\pm0.07$ & $0.395\pm0.011$  & 0.17 & 0.415 \\
J1824$-$2452A & -- & -- & -- & -- & 3.64 & 3.042 \\
J1857$+$0943 & $0.5\pm0.3$ & --& $0.15^{+0.2}_{-0.07}$ & $1.2^{+0.7}_{-0.4}$  & 0.70 & 1.168 \\
J1909$-$3744 & $0.810\pm0.003$ & $1.140\pm0.012$& $0.810\pm0.03$ & $1.23\pm0.05$  & 0.55 & 0.457 \\
J1939$+$2134 & $0.52\pm0.16$ & --& $0.40\pm0.16$ & $1.5^{+0.5}_{-0.3}$  & 3.58 & 3.550 \\
J2124$-$3358 & $2.4\pm0.4$ & --& $2.15\pm0.4$ & $1.39^{+0.08}_{-0.06}$  & 0.25 & 0.268 \\
J2129$-$5721 & -- & $3.2\pm1.5$& -- & --  & $>$2.55 & 1.686 \\
J2145$-$0750 & $1.84\pm0.17$ & --& $1.80\pm0.17$ & $0.53^{+0.06}_{-0.05}$  & 0.50 & 0.566 \\
\hline
\end{tabular}
\label{tab:Distances}
\end{center}
\end{table*}

The final parameters for the seven solitary pulsars J0711$-$6830, J1024$-$0719,  J1730$-$2304,  J1744$-$1134, J1824$-$2452A, J1939+2134, and J2124$-$3358 are given in Table 4. The binary pulsars are separated by the binary model used to describe their orbit. Parameters for the small-eccentricity pulsars described by the ELL1 model, PSRs J0613$-$0200, J1045$-$4509, J1603$-$7202, J1732$-$5049, J1857+0943,  J2129$-$5721, and J2145$-$0750 are presented in Table 5; DD model pulsars J1022+1001, J1600-3053, and J1643$-$1224 are presented in Table 6; and T2 model pulsars J0437$-$4715, J1713+0747, and J1909$-$3744 are presented in Table 7.

\subsection{PSR J0437$-$4715}

PSR J0437$-$4715 is the closest MSP currently known and the brightest at radio wavelengths. Van Straten et al. (2001) presented a timing solution including the full three-dimensional geometry of the binary orbit. An updated model was presented by V08.  This included a precise distance estimate derived from an orbital period-derivative measurement. 

\begin{figure*}
\centering
\includegraphics[trim=100 50 100 50,clip,width=\textwidth]{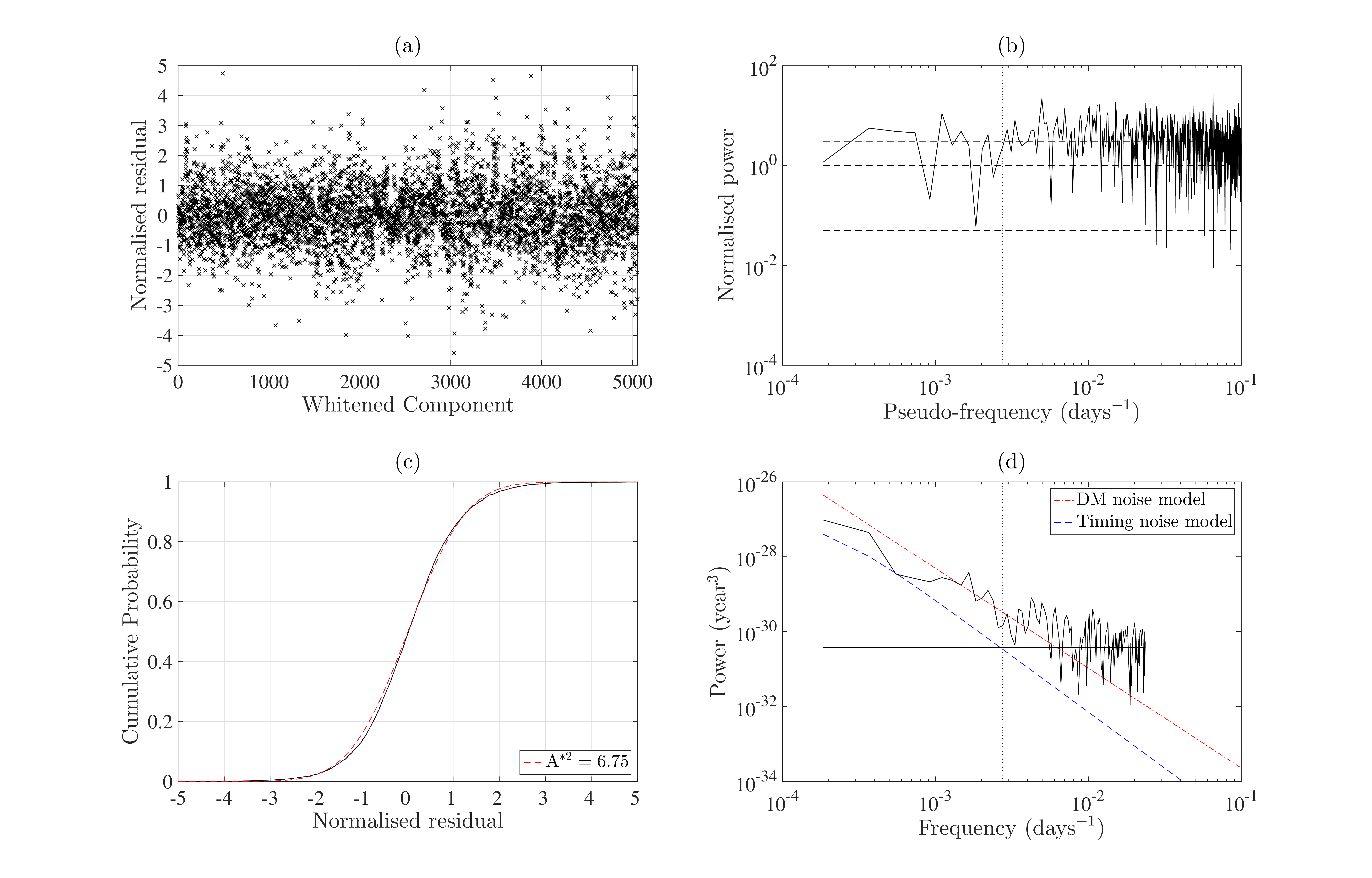}
\caption{(a) Whitened and normalised post-fit residuals for PSR J0437$-$4715. (b) Power spectra of whitened and normalised post-fit residuals. Dotted line marks $f=1$\,yr$^{-1}$ and dashed lines indicate expected mean and variance for the spectrum. The pseudo-frequency is determined by converting the whitened components to a pseudo-time-series using the ToAs of the unwhitened residuals.  (c) Cumulative distribution of whitened and normalised post-fit residuals (solid line) with expected distribution based on normal distribution with zero mean and unit variance (dashed line). Modified AD statistic for this distribution with the expected distribution is labelled. (d) Power spectra of post-fit residuals. Dotted line marks $f=1$\,yr$^{-1}$ and solid flat line is an estimate of the white-noise level. The timing noise model applies over entire data set in combined noise model, while the Kolmogorov DM noise model only applies to residuals prior to MJD 53430.}
\label{fig:J0437Analysis}
\end{figure*}

In Figure~\ref{fig:J0437Analysis}, we show the two components of the red-noise model with the power spectrum of the post-fit residuals (panel d).  As shown in panel (b), the model successfully whitens the residuals. Significant uncorrected DM noise is present in the early data.  Therefore, as expected, the timing noise model underestimates the total noise, whereas the DM noise model alone overestimates the total noise (since it does not apply to the entire dataset). There is excess noise at all frequencies (the mean of the power spectrum of normalised residuals is $>1$). This could result, for example, by additional uncorrected short-timescale correlated noise in the residuals. The normalised residuals do not pass the AD test ($A^{*2} = 6.75$). This may be because of instrumental effects, or because of pulse jitter (Shannon et al. 2014). Non-Gaussianity has been detected previously for PSR J0437$-$4715 by Lentati et al. (2014b). However it was shown that, at this level, the non-Gaussianity and high mean spectral power do not significantly affect the parameter measurements or uncertainties.

Using our new, precise measurement of $\dot{P_b}$, we can calculate an improved distance to PSR~J0437$-$4715 of $D=156.79 \pm 0.25$\,pc. We discuss this measurement in detail in Section 6.2. As shown in Table \ref{tab:Distances}, this measurement is consistent with independent distances, including our parallax distance measurement of $D=157 \pm 2$\,pc. V08 measured a high pulsar mass for this pulsar, of $M_p=1.76\pm0.2$\,M$_{\odot}$. Our improved measurement of the Shapiro delay reduces the uncertainty on the pulsar mass and we find $M_p=1.44\pm0.07$\,M$_{\odot}$, significantly smaller at the 1.5-$\sigma$ level. This improved mass measurement will be important for the NICER mission which will attempt to measure the neutron-star radius, probing the neutron star equation of state.

Kinematic contributions to the measured $\dot{\omega}$ are included in the timing model through the measurement of Kopeikin terms. We therefore expect that our $\dot{\omega}$ measurement is solely due to the effects of GR, as was reported in V08. Under this assumption, the reported $\dot{\omega}$ corresponds to a combined pulsar and companion mass of $M_p + M_c = 1.44\pm0.2$\,M$_{\odot}$, which is just consistent with the measured masses at the 1-$\sigma$ level. In analysis of future data sets for this pulsar, the improving  $\dot{\omega}$ measurement will be able to be used in combination with the Shapiro delay measurement to further constrain the pulsar mass.

\subsection{PSR J0613$-$0200}

The only new parameters included in our PSR~J0613$-$0200 timing model are those describing the sinusoidal annual variations in DM that were presented in Section 4.1. 

\subsection{PSR J0711$-$6830}

The pulsar timing model for this pulsar contains the same parameters as in V09.  As expected, our measurements are more precise than previous work.  We have not been able to obtain a parallax measurement for this pulsar because of its proximity to the ecliptic pole.

\subsection{PSR J1022$+$1001}

This pulsar has the smallest ecliptic latitude ($-0.06$\,deg) of all pulsars in the sample. Observations when the line-of-sight to the pulsar passes close to the Sun have been used by You et al. (2007) to study the solar corona.  However, these observations are generally removed for high-precision timing applications.  For our timing solutions, we removed the ToAs that were obtained when the line-of-sight to the pulsar passed within 5$^{\circ}$ of the Sun. The removal of these residuals produces normalised residuals that are consistent with zero mean and unit variance.

van Straten (2013) analysed 7.2 years of data for this pulsar from the Parkes radio telescope using a new method of polarimetric calibration to improve timing precision. The DR1 and DR1E datasets did not include this new calibration procedure. With our dataset we have detected a significant Shapiro delay using the traditional parameters $M_c$ and $\sin i$. However while we found $\sin i = 0.69\pm 0.18$, we did not measure a significant companion mass, $M_c=2.2\pm2.4$\,M$_{\odot}$. Our parameters are consistent with those of van Straten (2013).

We measured the orbital period-derivative, $\dot{P}_b = (5.5 \pm 2.3)\times 10^{-13}$, for the first time (this parameter was not measured by van Straten 2013). Assuming a pulsar mass of $M_p=1.4$\,M$_{\odot}$ and companion mass of $M_c=0.2$\,M$_{\odot}$, the expected $\dot{P}_b$ contribution from quadrupolar GW emission is $\dot{P}^{\rm GR}_{b}=-1.7\times 10^{-16}$, three orders of magnitude smaller than this measurement. We therefore expect that our measurement is an apparent orbital period increase caused by the Shklovskii effect. Given the proper motion of the pulsar, we can derive the distance to the pulsar, $D=1.5\pm0.5$\,kpc, which, as shown in Table~3, is consistent with other distance measurements.

The measured $\dot{x}$ is expected to be from the Kopeikin kinematic effects discussed in Section 3. From this, we place an upper limit on the inclination angle of the orbit of $i\leq 84$\,deg using the total proper motion presented in van Straten (2013), since we do not measure the proper motion in declination. The $\dot{\omega}$ measurement can have contributions from these same kinematic effects, but may also be consistent with the periastron advance expected from GR. Assuming that the observed $\dot{\omega}$ comes entirely from GR, we derive the combined mass of the system to be $M_p + M_c = 2.5\pm1.3$\,M$_{\odot}$, which is consistent with a neutron star -- white dwarf binary system.

\subsection{PSR J1024$-$0719}

The red timing noise for PSR J1024$-$0719 has a large spectral exponent (i.e., it is very steep). Our usual procedure, as described earlier, requires that we extrapolate the red-noise model obtained from the recent, multi-wavelength data into the earlier data. For PSR J1024$-$0719, we found that the red-noise model obtained from the recent data did not extrapolate well. Since DM noise is not detectable in the detrended data, we simply modelled the red noise in the entire dataset and did not use an additional DM noise model.

Our measurements are consistent with those of V09 and the measurement precision is improved in all cases. We measure a parallax of $\pi = 0.5\pm0.3$\,mas (prior to Lutz-Kelker bias correction), which was undetected by V09. A parallax was previously measured by Hotan et al. (2006) to be $\pi = 1.9\pm0.8$\,mas, which is also consistent with our measurements. However, since a parallax was undetected by V09, and PSR J1024$-$0719 has steep red noise that was unaccounted for by Hotan et al. (2006), the uncertainty for their original parallax measurement is likely to be severely underestimated.

\subsection{PSR J1045$-$4509}

A parallax measurement was presented by V09 of $\pi = 3\pm4$\,mas, but this measurement may have been affected by uncorrected red noise. We now make the first significant parallax measurement of $\pi = 2.2\pm1.1$\,mas.

\subsection{PSR J1600$-$3053}
We present the first significant measurement of $\dot{x}=(-4.2\pm0.7)\times 10^{-15}$, which is attributed to the proper motion of the system (Kopeikin 1996), and gives us an upper limit on the inclination angle of  $i\leq 67$\,deg. We detect the first significant parallax of $\pi = 0.48\pm0.11$\,mas (compared with $\pi = 0.2\pm0.3$\,mas given by V09). From the Shapiro delay companion mass measurement $M_c=0.34\pm 0.15$\,M$_{\odot}$ and the mass function, we can provide a constraint on the pulsar's mass: $M_p=2.4\pm 1.7$\,M$_{\odot}$.

\subsection{PSR J1603$-$7202}
For this pulsar, we have the first measurement of a parallax $\pi = 1.1\pm0.8$\,mas (prior to Lutz-Kelker bias correction). We also present the measured first derivatives of the orbital period, $\dot{P_b}=(3.1\pm1.5)\times10^{-13}$ and the projected semi-major axis, $\dot{x}=(1.36\pm0.16)\times 10^{-14}$. Since the GR contribution (for both $\dot{P_b}$ and $\dot{x}$) is negligible for this system, we use the proper motion of the pulsar and our $\dot{P}_b$ measurement to derive the distance to the pulsar, $D=3.9\pm1.8$\,kpc, which is marginally consistent with other distance measurements (Table 3). However all distance measurements for this pulsar are poor. Using the $\dot{x}$ measurement, we place an upper limit on the orbital inclination angle of $i\leq 31$\,deg. 

\subsection{PSR J1643$-$1224}
We present the first measurement of $\dot{\omega} = -0.007\pm 0.004$\,deg. An improved measurement of $\dot{x} = (-5.25\pm0.16) \times 10^{-14}$ allows us to derive an upper limit on the inclination angle of $i\leq 28$\,deg. For the measured $\dot{\omega}$ to be the result of GR effects, rather than the assumed kinematic effects, the combined mass of the system would need to be $M_p + M_c = 54$\,M$_{\odot}$; an order of magnitude larger than expected. We therefore expect that this measurement is not contaminated by GR contributions and instead results from kinematic effects. However since these measurements are not well determined and we do not detect a Shapiro delay, we are unable to find a unique solution for the Kopeikin terms $i$ and $\Omega$ and we therefore do not re-parametrise the orbit.

\subsection{PSR J1713$+$0747}

Splaver et al. (2005) reported on 12 years of timing observations of this pulsar from the Arecibo observatory.  Their analysis was carried out with the JPL DE405 solar system ephemeris and the time reference was TT(BIPM03). They obtained $\pi = 0.89 \pm 0.08$\,mas,  $\mu_\alpha \cos \delta = 4.917\pm 0.004$\,mas\,yr$^{-1}$, and $\mu_\delta = -3.933\pm 0.01$\,mas\,yr$^{-1}$. The orbital projection effects caused by this proper motion allowed them to determine $\Omega = (87\pm 6)^\circ$.  Their analysis was carried out  by ``whitening" the residuals using eight time derivatives of the pulse frequency.  C11 showed that such whitening can lead to underestimated parameter uncertainties.

V09 obtained a less precise parallax determination of $\pi = 0.94\pm 0.10$\,mas and a proper motion components of $\mu_\alpha \cos \delta = 4.924 \pm 0.10$\,mas\,yr$^{-1}$ and $\mu_\delta = -3.85 \pm 0.02$\,mas\,yr$^{-1}$. They obtained $i = (78.6\pm 1.7)^\circ$ and $\Omega = (67\pm 17)^\circ$. V09 also included $\dot{P}_b = (41\pm 20)\times 10^{-13}$.

The most recent VLBI observations of this pulsar (Chatterjee et al. 2009) give proper motion components of $\mu_\alpha = 4.75^{+0.16}_{-0.07}$\,mas\,yr$^{-1}$ and $\mu_\delta = -3.67^{+0.06}_{-0.15}$\,mas\,yr$^{-1}$ and parallax of $\pi = 0.95^{+0.06}_{-0.05}$\,mas. These values are in fair agreement with our values in Table 7.

Zhu et al. (2015) have analysed 21 years of timing data from the North American Nanohertz Observatory for Gravitational Waves (NANOGrav) for this pulsar to conduct tests of theories of gravity. In their analysis they measured parameters using a number of different noise models. Using \textsc{Tempo2} with a jitter-based white-noise model and a red-noise model, they measured $\pi = 0.87\pm 0.03$\,mas, $\mu_\alpha \cos \delta = 4.915 \pm 0.003$\,mas\,yr$^{-1}$, and $\mu_\delta = -3.914 \pm 0.005$\,mas\,yr$^{-1}$. For the binary model, they measured the Kopeikin terms $i = (71.9\pm 0.7)^\circ$, and $\Omega = (88\pm 2)^\circ$, as well as a companion mass of $M_c=0.286\pm 0.012$\,M$_{\odot}$ and $\dot{P_b}=(0.36\pm0.17)\times 10^{-12}$. These parameters are consistent with, and more precise than our measurements below because of the longer data span and higher timing precision of the NANOGrav dataset for this pulsar.

With our analysis we obtain $\dot{P_b}=(1.7\pm0.7)\times 10^{-12}$. The intrinsic $\dot{P}^{\rm GR}_{b}$ from GW emission is negligible and so we expect that this result comes from the Shklovskii effect.  This provides a pulsar distance of $D=3.1\pm1.2$\,kpc, which is marginally consistent with other distance measurements (Table 3). We also use the Shapiro delay companion mass measurement $M_c=0.32\pm 0.05$\,M$_{\odot}$ and the mass function, to calculate the pulsar mass, $M_p=1.7\pm 0.4$\,M$_{\odot}$.

\subsection{PSR J1730$-$2304}

We present the first measurement of a parallax for this pulsar of $\pi = 1.5\pm0.3$\,mas, prior to Lutz-Kelker bias correction. All other parameters are consistent with the previous values from V09.

\subsection{PSR J1732$-$5049}

V09 was only able to determine $\mu_\delta = -9.3\pm 0.7$\,mas\,yr$^{-1}$. We now also present a measurement of the proper motion in right ascension, $\mu_{\alpha} \cos \delta = -0.41\pm 0.09$\,mas\,yr$^{-1}$, but parallax was not detected.

\subsection{PSR J1744$-$1134}

All parameters are consistent with V09 after accounting for apparent changes resulting from the different solar-system ephemeris and time standard used in the analysis. 


\subsection{PSR J1824$-$2452A}

PSR J1824$-$2452A is a solitary pulsar located in the globular cluster M28. The timing residuals for this pulsar exhibit red noise, which may be caused by acceleration within the cluster potential, or timing noise. Our timing model includes dDM/dt but, as we do not include any single-frequency data for this pulsar, we use the original Cholesky routines with a single noise model. We did not measure a significant proper motion in declination nor a parallax even though V09 did publish a proper motion in declination.

\subsection{PSR J1857$+$0943}
This pulsar is in an orbit that is highly inclined to our line of sight, allowing for a precise measurement of the Shapiro delay. We do not improve parameter uncertainties for every parameter since V09 made use of publicly available data from the Arecibo observatory to extend the dataset, while we chose to use only the PPTA DR1E data set. Using our Shapiro delay companion mass measurement $M_c=0.25\pm 0.03$\,M$_{\odot}$ and the mass function, we calculate the pulsar mass to be $M_p=1.5\pm 0.2$\,M$_{\odot}$.

\subsection{PSR J1909$-$3744}
The narrow pulse width, particularly at 10\,cm, allows us to achieve very low ToA uncertainties. Recent PPTA data for this pulsar, timed to sub-100\,ns precision over more than ten years has led to the most stringent limit on the stochastic GW background to date (Shannon et al. 2015). For this reason, this pulsar is an important tool for testing models of galaxy and supermassive black hole formation.

For this pulsar, we use a corrected version of the DR1E dataset that is described in Shannon et al. (2013). Previously undetected instrumental offsets were found and corrected, and additional archival 50cm observations were included to allow measurement of $\Delta$DM over the entire dataset. We also include an additional jump corresponding to a software upgrade at MJD 55319.8 that was identified by Shannon et al. (2015). We include dDM/dt in the timing model, which removes the majority of the DM noise. While there is no evidence for red noise in the 10\,cm residuals (Shannon et al. 2015), we identify some slight red noise originating from the 20\,cm and 50\,cm residuals. This could be the result of instrumental noise or residual interstellar dispersion noise. Because of this noise, we included a red-noise model that sufficiently whitens the residuals.

V09 did not include Kopeikin terms in the timing model, but instead fitted for $\dot{x}$; $\dot{\omega}$ was not measured. We now include the Kopeikin terms, giving the inclination angle, $i=93.52 \pm 0.09$\,deg, and the longitude of ascending node, $\Omega=39 \pm 10$\,deg. We measure the orbital period-derivative to be $\dot{P}_b = (5.03 \pm 0.06)\times 10^{-13}$. The expected contribution from quadrupolar GW emission to this measurement is $\dot{P}^{\rm GR}_{b}=-2.7\times 10^{-15}$; two orders of magnitude smaller than this measurement. This expected value was calculated from the measured companion mass $M_c=0.2067\pm 0.0019$\,M$_{\odot}$ and the calculated pulsar mass from the Shapiro delay and mass function of $M_p=1.47\pm 0.03$\,M$_{\odot}$. From this $\dot{P}_b$ measurement and the proper motion, we derive a distance of $D=1.140\pm0.012$\,kpc, which is consistent with the parallax distance. The distances derived from the DM and galactic electron density models are evidently under-estimated.

\subsection{PSR J1939$+$2134}
PSR J1939$+$2134 was the first MSP discovered (Backer et al. 1982), and it is currently the second fastest spinning pulsar known. The timing residuals for this pulsar are dominated by red noise.

\subsection{PSR J2124$-$3358}

All measured parameters for this pulsar are consistent with V09 and the uncertainties have decreased in all cases.

\subsection{PSR J2129$-$5721}
We have the first measurement of the orbital period-derivative, $\dot{P_b}=(7.9\pm3.6)\times10^{-13}$. Using this measurement and the proper motion, we derive a distance of $D=3.2\pm1.5$\,kpc, which is consistent with the DM distances (Table 3), however all distance measurements for this pulsar are poor. We do not yet detect a parallax for this pulsar because of its proximity to the ecliptic pole.

\subsection{PSR J2145$-$0750}
We have the first measurement of $\dot{x}=(8.0\pm0.8)\times10^{-15}$, resulting from the proper motion of the pulsar. Using this value, we place an upper limit on the inclination angle of the orbit of $i\leq 69$\,degrees. All parameters are consistent with the previous results. V09 published $\dot{x}$, $\dot{P_b}$, and $\dot{\omega}$ values, but the measurements were not significant. Furthermore, the uncertainties were likely to be underestimated because of the red noise present in these observations.

\section{Discussion}

\subsection{Advantages of using the split-Cholesky algorithm} 

The methodology that we have used in this paper is based on traditional, frequentist analysis of pulsar timing residuals. An alternate approach is through Bayesian algorithms, such as those described in van Haasteren et al. (2009), van Haasteren \& Levin (2013), or the \textsc{Temponest} algorithm developed by Lentati et al. (2014a). These algorithms have successfully been used by the NANOGrav and European Pulsar Timing Array (EPTA) groups (e.g. van Haasteren et al. 2011, Arzoumanian et al. 2014). When these same algorithms are applied to PPTA datasets, uncertainties arise in the noise models because of the covariance between DM variations and timing noise processes in the single-wavelength (20\,cm) early data. At present, there is no way for current implementations of the Bayesian algorithms to model such non-stationary red noise in the way that the split-Cholesky algorithm allows. Instead the Bayesian algorithms assume that the noise is wide-sense stationary. Constructing separate red-noise models for the timing noise and DM variations allows us to better understand our noise model, by avoiding large uncertainties in our early data. We therefore chose to use a frequentist approach to analyse our dataset since it is less computationally expensive than the Bayesian alternatives and gives us greater control of our noise models.

\begin{figure}
\centering
\includegraphics[trim=0 50 0 0,clip,width=21pc]{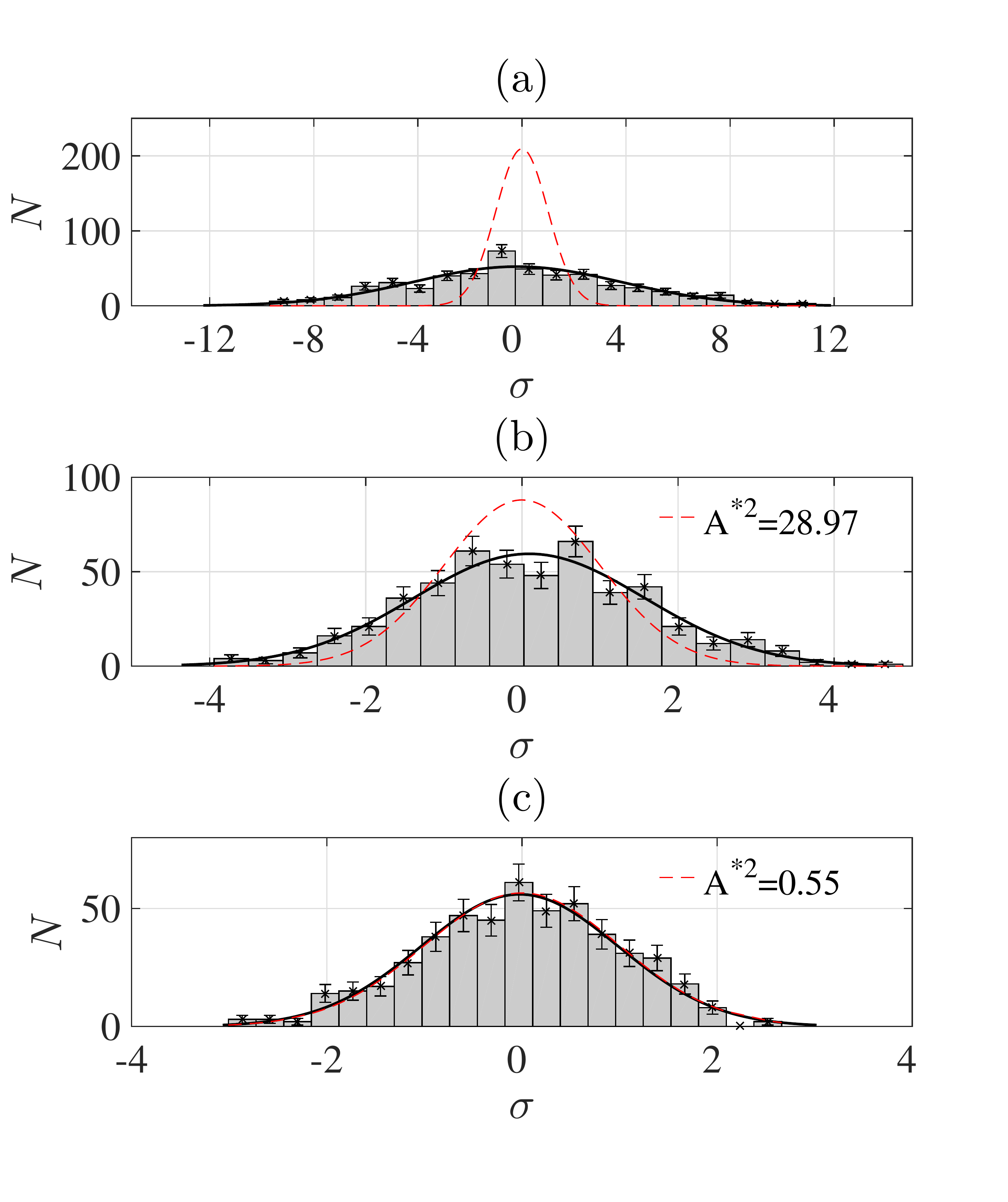}
\caption{Distribution of parallax values from timing model fits to 500 realisations of simulated PSR J0437$-$4715 data. $\sigma$ is the number of standard deviations from the true value for each of the $N$ realisations. We used three different noise treatments: (a) No red-noise model, (b) red-noise model from DR1 data, extrapolated to apply over the entire dataset was used with the Cholesky algorithm, and (c) two-component red-noise model with Kolmogorov DM model is used in split-Cholesky algorithm. For each panel, the black line is a normal distribution fit to the distribution and the red, dashed line is a normal distribution with zero mean and standard deviation equal to the average of the standard \textsc{tempo2} uncertainties for the 500 realisations, scaled to the same area as the black-line distribution.}
\label{fig:PXComparison}
\end{figure}

\begin{figure*}
\centering
\includegraphics[trim=100 270 200 50,clip,width=45pc]{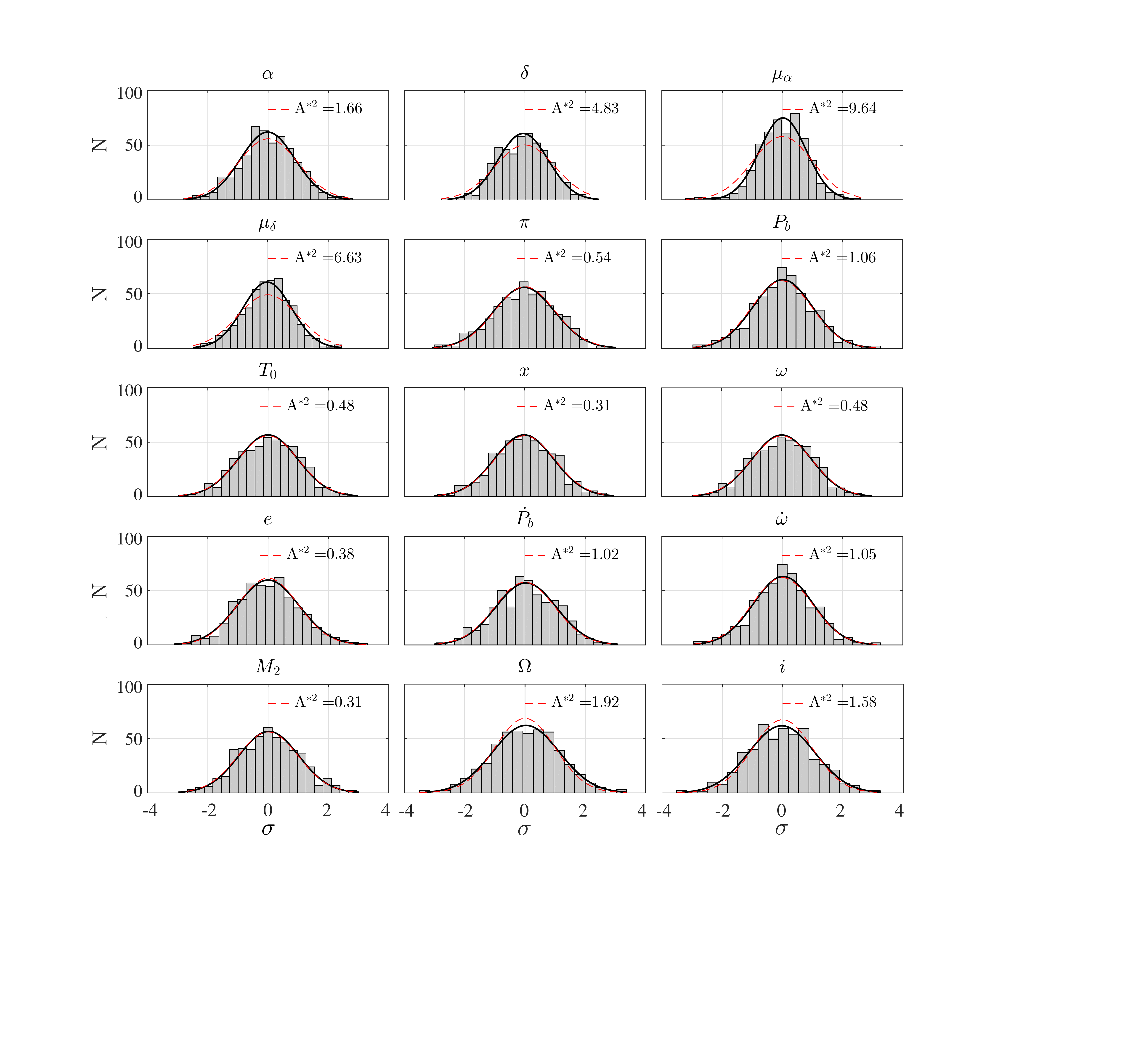}
\caption{Distributions of parameter values from timing model fits with the split-Cholesky algorithm to 500 realisations of simulated PSR J0437$-$4715 data. $\sigma$ is the number of standard deviations from the true value for each realisation. Data were simulated with statistically identical red noise properties to those of PSR J0437$-$4715.}
\label{fig:splitCholeskyResults}
\end{figure*}

To demonstrate the necessity for the split-Cholesky algorithm with our dataset, we created 500 realisations of PSR J0437$-$4715 data with red noise and DM noise at the level presented in Section 4. The parameters in the timing model were fitted (including jumps and DM variations) with three different noise treatments; no red-noise model, extrapolated DR1 red-noise model in the Cholesky algorithm only, and a two-component red-noise model with Kolmogorov DM noise in the split-Cholesky algorithm. In Figure \ref{fig:PXComparison} we show the distribution of post-fit parallax values represented by the number of standard deviations from the true value. When no noise model is used (panel a), the parameter uncertainties are clearly underestimated. When the single-component red-noise model is used (panel b), there is a significant improvement but the parameter uncertainties remain underestimated. Finally, when a two-component red-noise model is used with the split-Cholesky algorithm (panel c), we can accurately model the total red noise for the pulsar, and as a result we avoid underestimation of uncertainties. This is true for all parameters with the exception of $\nu$ and $\dot{\nu}$ (as was the case in the original Cholesky algorithm; see C11). The distributions of post-fit values for each parameter in the PSR J0437$-$4715 timing model (excluding $\nu$ and $\dot{\nu}$), using the split-Cholesky method, are given in Figure \ref{fig:splitCholeskyResults} with the Anderson-Darling statistic used to test the distribution. We see that the distributions are consistent with the expected zero mean, unit variance distribution for all parameters except for declination and proper motion, which have slightly overestimated uncertainties. 

\subsection{PSR J0437$-$4715 Kinematic distance measurement from $\mathbf{\dot{P}_{b}}$}

As described in Section 3, the measurement of $\dot{P_{b}}$ for PSR J0437$-$4715 can be used to measure the pulsar's distance, using Equation 2. Contributions to the observed $\dot{P_{b}}$ value can come from changes intrinsic to the pulsar system, $\dot{P}_{b}^{\rm{int}}$, from the kinematic or Shklovskii effect, $\dot{P}_{b}^{\rm{kin}}$, and from differential acceleration of the solar system and pulsar system caused by mass in the Galaxy, $\dot{P}_{b}^{\rm{Gal}}$.  Therefore,
\begin{equation}
\dot{P}_{b}^{\rm{obs}} = \dot{P}_{b}^{\rm{int}} + \dot{P}_{b}^{\rm{Gal}} + \dot{P}_{b}^{\rm{kin}}
\end{equation}
where we have measured $\dot{P}_{b}^{\rm{obs}}=\left(3.7276 \pm 0.0058\right) \times 10^{-12}$.

The intrinsic orbital decay for neutron star-white dwarf systems such as PSR J0437$-$4715 is dominated by quadrupolar GW emission, which can be calculated using the relation:
\begin{dmath}
{\dot{P}_{b}^{\rm{int}}\simeq\dot{P}_{b}^{\rm{GR}}=-\frac{192\pi G^{5/3}}{5c^5}\left(P_b/2\pi\right)^{-5/3}\left(1-e^2\right)^{-7/2}}\times\left(1+\frac{73}{24}e^2+\frac{37}{96}e^4\right)m_p m_c\left(m_p+m_c\right)^{-1/3}
\end{dmath}
(Peters \& Mathews 1963, Taylor \& Weisberg 1982). For PSR J0437$-$4715 this GR contribution is $\dot{P}_{b}^{\rm{GR}}=-3.2\times 10^{-16}$, which is smaller than the value calculated by V08 because of their measurement of a high pulsar mass. This GR contribution is an order of magnitude smaller than the uncertainty in the observed value.

The galactic acceleration component can be estimated by combining the accelerations from differential rotation and the galactic potential. The differential rotation acceleration is found using the galactic longitude and distance to the pulsar, and the galactocentric distance and circular velocity of the Sun. Acceleration in the the galactic potential varies as a function of height and can be computed from a model of the local surface density of the Galaxy given the galactic latitude of the pulsar and its parallax distance. V08 used the Holmberg \& Flynn (2004) and calculate the total galactic contribution to be $\dot{P}_{b}^{\rm{Gal}}=\left(-1.8 -0.5\right) \times 10^{-14}=-2.3\times 10^{-14}$. Since the value of the galactic acceleration is larger than the uncertainty of our observed value, the uncertainty in this component may become important. Bovy et al. (2012) find that the uncertainty for the circular velocity of local sources is approximately 3\% by using data from the Apache Point Observatory galactic Evolution Experiment. From Holmberg \& Flynn (2004), the uncertainty in the surface density resulting in the vertical component of acceleration is approximately 10\%. Using these uncertainties, we have $\dot{P}_{b}^{\rm{Gal}}=\left(-2.3\pm 0.08\right)\times 10^{-14}$, which is small compared to the uncertainty in our measurement.

The kinematic contribution to $\dot{P}_{b}^{\rm{obs}}$ because of the Shklovskii effect and the distance to the pulsar can be found through
\begin{equation}
\dot{P}_{b}^{\rm{obs}}-\dot{P}_{b}^{\rm{GR}}-\dot{P}_{b}^{\rm{Gal}}\simeq\dot{P}_{b}^{\rm{kin}}=\frac{\mu^{2}D}{c}P_b
\end{equation}
where $\mu$ is the total proper motion of the pulsar and $D$ is the distance. This gives $\dot{P}_{b}^{\rm{kin}}=\left(3.7513\pm 0.006 \right)\times 10^{-12}$ and $D=156.79 \pm 0.25$\, pc, which is consistent with our independently measured parallax distance of $D=156.9 \pm 2.2$\,pc. This is the most precise distance measurement for all pulsars and one of the most fractionally precise distance measurements for any star. Our distance measurement is also consistent with the VLBI parallax distance of $D=156.3 \pm 1.3$\,pc, measured by Deller et al. (2008). Since $\dot{P}_{b}$ produces an effect in the residuals that grows over time, we can expect the uncertainty in $\dot{P}_{b}^{\rm{obs}}$ to decrease significantly in future data sets, leaving the distance uncertainty to be dominated by the uncertainty in models used to calculate the contributions from galactic acceleration. Precise distance measurements such as this will be important to PTAs in the future since it allows the use of the pulsar term in single-source GW detection, which is essential for accurately determining the source location (Lee et al. 2011).

\section{Conclusion}

We have presented new models for red noise caused by DM variations in single-frequency data sets, obtained by first including deterministic components in the timing model, and then modelling the covariance function of the remaining noise. For 14 of the pulsars in our sample, we detect a linear trend in the DM variations and include this in the timing model, and for four pulsars (PSRs J0613$-$0200, J1045$-$4509, J1643$-$1224, and J1939+2134) we also include new parameters that describe annual DM variations. The new DM noise models apply only to the early data where excess noise is present, and are used in conjunction with a red-noise model for the frequency-independent noise that is present in the entire dataset. These two-component models were used in the new ``split-Cholesky" algorithm to whiten the residuals to provide unbiased parameter measurements. We have described this algorithm and demonstrated its effectiveness on simulated data. Model parameters were shown to be unbiased and have accurate uncertainties through simulations based on PSR J0437$-$4715. 

Determining new timing models for the 20 PPTA pulsars in the DR1E dataset required these new models and algorithm because of the non-stationary red noise for most pulsars. The models we present provide the best description of the noise currently possible with the PPTA data sets, and result in the most accurate and precise parameter measurements to date for most pulsars in our sample, as well as the detection of several new parameters. Most notably, we presented the first significant parallax measurements for PSRs J1024$-$0719, J1045$-$4509, J1600$-$3053, J1603$-$7202, and J1730$-$2304, and determined the distance to PSR J0437$-$4715 at the 0.16\% level of precision. We also measured an improved pulsar mass for PSR J0437$-$4715, which at $M_p=1.44\pm0.07$\,M$_{\odot}$, is somewhat lower than the previous measurement.

The analysis described here can easily be applied to future PPTA data releases and to any PTA data with non-stationary noise processes. Longer datasets for PSR J0437$-$4715 will further improve the distance measurements based on both the parallax, and the orbital period-derivative from Schklovskii acceleration. If the timing parallax distance becomes more precise than the current VLBI distance, these two independent measurements can be used for example to further improve current constraints on the change to Newton's Gravitational constant (Freire et al. 2012). It may also be possible to measure, or place limits on, the acceleration of the pulsar system caused by mass in the Galaxy. With longer datasets, improving $\dot{\omega}$ measurements, particularly for PSR J0437$-$4715, can compliment the Shapiro delay to further improve measurements of the pulsar mass.

\section*{References}
Akaike H., 1998, Springer Series in Statistics, 199-213\\
Anderson T. W., Darling D. A., 1954, Journal of the \indent American Statistical Association, 49, 268\\
Arzoumanian Z. et al., 2014, ApJ, 794, 141\\
Backer D. C. et al., 1982, Nat, 300, 615\\
Bailes M. et al., 1994, ApJ, 425, L41\\
Bell J. F., Bailes M., 1996, ApJ, 456, L33\\
Bell J. F. et al., 1997, MNRAS, 286, 463\\
Bovy J. et al, 2012, ApJ, 759, 131\\ 
Champion D. J. et al., 2010, ApJ, 720, 201\\
Chatterjee S. et al., 2009, ApJ, 698, 250\\
Coles W. et al., 2011, MNRAS, 418, 561\\
Coles W. et al., 2015, ApJ, 808, 113\\
Cordes J. M., Lazio T. J. W., 2002, arXiv:astro-ph/0207156\\
Dai S. et al., 2015, MNRAS, 449, 3223\\
Damour T., Deruelle N., 1986, AIHS, 44, 263\\
Deller A. T. et al., 2008, ApJ, 685, L67\\ 
Demorest P. B. et al., 2013, ApJ, 762, 94\\
Edwards R. T. et al., 2001, MNRAS, 326, 358\\
Edwards R. T., Hobbs G. B., Manchester R. N., 2006, \indent MNRAS, 372, 1549\\
Fiedler R. L. et al., 1987, Nat, 326, 675\\
Freire P. C. C. et al., 2012, MNRAS, 423, 3328\\
Gendreau K. C., 2012, SPIE, 8443, 8\\
Hobbs G. B., Edwards R. T., Manchester R. N., 2006, \indent MNRAS, 369, 655\\
Hobbs G. B. et al., 2011, PASA, 28, 202\\
Hobbs G. B. et al., 2012, MNRAS, 427, 2780\\
Holmberg J., Flynn C., 2004, MNRAS, 352, 440\\
Hotan A. W., Bailes M., Ord S. M., 2006, MNRAS, 369, \indent 1502\\
Keith M. J. et al., 2013, MNRAS, 429, 216\\
Kopeikin S. M., 1995, ApJ, 439, L5\\
Kopeikin S. M., 1996, ApJ, 467, L93\\
Kramer M. et al., 2006, Sci, 314, 97\\
Kramer M., Champion D. J., 2013, CQG, 30, 4009\\
Lange Ch. et al., 2001, MNRAS, 326, 274\\ 
Lee K. J. et al., 2011, MNRAS, 414, 3251\\
Lentati L. et al., 2014, MNRAS, 437, 3004\\
Lentati L., Hobson M. P.,  Alexander P., 2014, MNRAS, 444, 3863\\
Lutz T. E.,  Kelker D. H., 1973, PASP, 85, 573\\
Maitia V., Lestrade J.-F., Cognard I., 2003, ApJ, 582, 972\\
Manchester R. N. et al., 2013, PASA, 30, 31\\
Ord S. M. et al., 2006, MNRAS, 371, 337\\
Os{\l}owski S. et al., 2011, MNRAS, 418, 1258\\
Peters P. C., Mathews J., 1963, PhysRev, 131, 435\\
Sandhu J. S., 1997, ApJ, 478, 95\\
Shannon R. M., Cordes J. M., 2010, ApJ, 725, 1607\\
Shannon R. M. et al., 2014, MNRAS, 443, 1463\\
Shannon R. M. et al., 2013, Science, 342, 334\\
Shannon R. M. et al., 2015, Science, 349, 1522\\
Shapiro I. I., 1964, PhysRev, 13, 789\\
Shklovskii I. S., 1970, SvA, 13, 562\\
Splaver E. M. et al., 2005, ApJ, 620, 405\\
Stephens M. A., 1974, Journal of the American Statistical \indent Association, 69, 730\\
Taylor J. H., Weisberg J. M., 1982, ApJ, 253, 908\\
Taylor J. H., Cordes J. M., 1993, ApJ, 411, 674\\
Thorsett S. E., Chakrabarty D., 1999, ApJ, 512, 288\\
Toscano M. et al., 1999, MNRAS, 307, 925\\
van Haasteren R. et al., 2009, MNRAS, 395, 1005\\
van Haasteren R. et al., 2011, MNRAS, 414, 3117\\
van Haasteren R., Levin Y., 2013, MNRAS, 428, 1147\\
van Straten W. et al., 2001, Nat, 412, 158\\
van Straten W., Bailes M., 2003. ASPC, 302, 65\\
van Straten W., 2013, ApJ, 204, 13\\
Verbiest J. P. W. et al., 2008, ApJ, 679, 675\\
Verbiest J. P. W. et al., 2009, MNRAS, 400, 951\\
Verbiest J. P. W. et al., 2012, ApJ, 755, 39\\
Wang J. B. et al., 2015, MNRAS, 446, 1657\\
You X. P. et al., 2007, MNRAS, 378, 493\\
Zhu W. W. et al., 2015, arXiv:1504.00662\\
Zhu X.-J. et al., 2014, MNRAS, 444, 3709\\

\section*{Acknowledgments}
We thank J. Verbiest for assistance with the use of the Lutz-Kelker Bias correction webpage, and the referee for their useful feedback. The Parkes radio telescope is part of the Australia Telescope National Facility which is funded by the Commonwealth of Australia for operation as a National Facility managed by CSIRO. YL and GH are recipients of ARC Future Fellowships (respectively, FT110100384 and FT120100595). YL, MB, WvS, and PDL are supported by ARC Discovery Project DP140102578. SO is supported by the Alexander von Humboldt Foundation. VR is a recipient of a John Stocker postgraduate scholarship from the Science and Industry Endowment Fund of Australia. LW and XZ acknowledge support from the Australian Research Council. J-BW is supported by NSFC project No.11403086 and West Light Foundation of CAS XBBS201322. XPY is supported by NSFC project U1231120, FRFCU project XDJK2015B012 and China Scholarship Council (CSC).

\section*{Appendix}
The pulsar parameters are listed in the tables below, and a discussion for each pulsar is given in Section 5. The model parameter files, ToA files, and red-noise models for each pulsar are available from the CSIRO Data Access Portal: http://dx.doi.org/10.4225/08/561EFD72D0409.

\clearpage 
\pagestyle{empty}
\oddsidemargin=1.4cm
\evensidemargin=1.4cm
\topmargin=5.2cm
\begin{landscape}
\begin{table*}
\scriptsize
\begin{adjustwidth}{-1cm}{}
\caption{Parameters for the solitary pulsars J0711$-$6830, J1024$-$0719,  J1730$-$2304,  J1744$-$1134, J1824$-$2452A, J1939+2134, and J2124$-$3358. Numbers in brackets are the \textsc{tempo2} 1-sigma uncertainties on the last quoted decimal place, including split-Cholesky analysis.}
\begin{tabular}{llllllll}
\hline															
\hline															
Pulsar name\dotfill 	&	 J0711$-$6830  	&	 J1024$-$0719  	&	 J1730$-$2304  	&	 J1744$-$1134  	&	 J1824$-$2452A  	&	 J1939+2134  	&	 J2124$-$3358  	\\
MJD range\dotfill 	&	 49373.6---55619.2  	&	 50117.5---55619.5  	&	 49421.9---55598.8  	&	 49729.1---55619.0  	&	 53518.8---55619.1  	&	 49956.5---55619.0  	&	 49489.9---55618.2  	\\
Data span (yr)\dotfill 	&	17.1	&	15.06	&	16.91	&	16.13	&	5.75	&	15.5	&	16.78	\\
Number of TOAs\dotfill 	&	566	&	493	&	390	&	534	&	313	&	397	&	652	\\
Rms timing residual ($\mu s$)\dotfill 	&	2.0	&	10.4	&	1.9	&	0.5	&	5.5	&	5.8	&	2.9	\\
\hline															
\multicolumn{2}{c}{Measured Quantities}  			&		&		&		&		&		&		\\
\hline															
Right ascension (RA), $\alpha$ (hh:mm:ss)\dotfill 	&	  07:11:54.189114(13) 	&	  10:24:38.678633(7) 	&	  17:30:21.66624(8) 	&	  17:44:29.4057891(11) 	&	  18:24:32.00788(3) 	&	  19:39:38.561213(3) 	&	  21:24:43.849372(10) 	\\
Declination (DEC), $\delta$ (dd:mm:ss)\dotfill 	&	 $-$68:30:47.41446(8) 	&	 $-$07:19:19.36778(19) 	&	 $-$23:04:31.19(2) 	&	 $-$11:34:54.68126(8) 	&	 $-$24:52:10.834(8) 	&	 +21:34:59.12628(6) 	&	 $-$33:58:44.8500(3) 	\\
Pulse frequency, $\nu$ (s$^{-1}$)\dotfill 	&	 182.1172346685786(10) 	&	 193.71568347859(4) 	&	 123.1102871605879(5) 	&	 245.4261197130557(4) 	&	 327.40559048006(3) 	&	 641.92822645342(4) 	&	 202.793893782574(6) 	\\
First derivative of pulse frequency, $\dot{\nu}$ (s$^{-2}$)\dotfill 	&	 $-$4.94405(11)$\times 10^{-16}$ 	&	 $-$6.9523(18)$\times 10^{-16}$ 	&	 $-$3.05917(4)$\times 10^{-16}$ 	&	 $-$5.38173(3)$\times 10^{-16}$ 	&	 $-$1.735302(7)$\times 10^{-13}$ 	&	 $-$4.33106(3)$\times 10^{-14}$ 	&	 $-$8.4597(3)$\times 10^{-16}$ 	\\
Proper motion in RA, $\mu_{\alpha} \cos \delta$ (mas\,yr$^{-1}$)\dotfill 	&	 $-$15.57(3) 	&	 $-$35.33(4) 	&	 20.264(19) 	&	 18.790(6) 	&	 $-$0.69(13) 	&	 0.087(16) 	&	 $-$14.14(4) 	\\
Proper motion in DEC, $\mu_{\delta}$ (mas\,yr$^{-1}$)\dotfill 	&	 14.24(3) 	&	 $-$48.32(8) 	&	--	&	 $-$9.40(3) 	&	--	&	 $-$0.41(3) 	&	 $-$50.08(9) 	\\
Parallax, $\pi$ (mas)\dotfill 	&	--	&	 0.5(3) 	&	 1.5(3) 	&	 2.53(7) 	&	--	&	 0.52(16) 	&	 2.4(4) 	\\
\hline															
\multicolumn{2}{c}{Set Quantities}  			&		&		&		&		&		&		\\
\hline															
Dispersion measure, DM (cm$^{-3}$pc)\dotfill 	&	18.4099	&	6.48803	&	9.61634	&	3.13695	&	119.892	&	71.0227	&	4.60096	\\
\hline															
\multicolumn{2}{c}{Derived Quantities} 			&		&		&		&		&		&		\\
\hline															
$\log_{10}$(Characteristic age, yr) \dotfill 	&	9.77	&	9.64	&	9.8	&	9.86	&	7.48	&	8.37	&	9.58	\\
$\log_{10}$(Surface magnetic field strength, G) \dotfill 	&	8.46	&	8.5	&	8.61	&	8.29	&	9.35	&	8.61	&	8.51	\\
$\log_{10}$(Edot, ergs/s) \dotfill 	&	33.55	&	33.73	&	33.17	&	33.72	&	36.35	&	36.04	&	33.83	\\
\hline															
\end{tabular}
\end{adjustwidth}
\end{table*}
\end{landscape}

\clearpage 
\pagestyle{empty}
\oddsidemargin=1.4cm
\evensidemargin=1.4cm
\topmargin=5.2cm
\begin{landscape}
\begin{table*}
\begin{adjustwidth}{-1cm}{}
\scriptsize
\caption{Parameters for the binary pulsars described by the small-eccentricity ELL1 binary model, PSRs J0613$-$0200, J1045$-$4509,  J1603$-$7202,  J1732$-$5049, J1857+0943, J2129$-$5721, and J2145$-$0750. Numbers in brackets are the \textsc{tempo2} 1-sigma uncertainties on the last quoted decimal place, including split-Cholesky analysis.}
\begin{tabular}{llllllll}
\hline															
\hline															
Pulsar name\dotfill 	&	 J0613$-$0200  	&	 J1045$-$4509  	&	 J1603$-$7202  	&	 J1732$-$5049  	&	 J1857+0943  	&	 J2129$-$5721  	&	 J2145$-$0750  	\\
MJD range\dotfill 	&	 51526.6---55619.3  	&	 49405.5---55619.5  	&	 50026.1---55618.8  	&	 52647.1---55582.2  	&	 53086.9---55619.0  	&	 49987.4---55618.3  	&	 49517.8---55618.2  	\\
Data span (yr)\dotfill 	&	11.21	&	17.01	&	15.31	&	8.04	&	6.93	&	15.42	&	16.7	\\
Number of TOAs\dotfill 	&	639	&	646	&	493	&	244	&	291	&	448	&	972	\\
Rms timing residual ($\mu s$)\dotfill 	&	1.0	&	11.9	&	2.7	&	3.1	&	1.1	&	1.4	&	1.7	\\
\hline															
\multicolumn{2}{c}{Measured Quantities}  			&		&		&		&		&		&		\\
\hline															
Right ascension (RA), $\alpha$ (hh:mm:ss)\dotfill 	&	  06:13:43.975503(4) 	&	  10:45:50.18696(3) 	&	  16:03:35.67751(4) 	&	  17:32:47.766731(19) 	&	  18:57:36.390848(4) 	&	  21:29:22.766966(12) 	&	  21:45:50.46148(3) 	\\
Declination (DEC), $\delta$ (dd:mm:ss)\dotfill 	&	 $-$02:00:47.21147(14) 	&	 $-$45:09:54.1223(4) 	&	 $-$72:02:32.72985(19) 	&	 $-$50:49:00.1917(4) 	&	 +09:43:17.21458(9) 	&	 $-$57:21:14.21183(12) 	&	 $-$07:50:18.4759(12) 	\\
Pulse frequency, $\nu$ (s$^{-1}$)\dotfill 	&	 326.6005620676858(16) 	&	 133.793149554823(14) 	&	 67.376581131844(3) 	&	 188.233512213289(4) 	&	 186.4940784047797(7) 	&	 268.3592273587338(16) 	&	 62.2958878423832(13) 	\\
First derivative of pulse frequency, $\dot{\nu}$ (s$^{-2}$)\dotfill 	&	 $-$1.02293(3)$\times 10^{-15}$ 	&	 $-$3.1616(8)$\times 10^{-16}$ 	&	 $-$7.0956(12)$\times 10^{-17}$ 	&	 $-$5.0296(6)$\times 10^{-16}$ 	&	 $-$6.20417(14)$\times 10^{-16}$ 	&	 $-$1.501784(14)$\times 10^{-15}$ 	&	 $-$1.15599(7)$\times 10^{-16}$ 	\\
Proper motion in RA, $\mu_{\alpha} \cos \delta$ (mas\,yr$^{-1}$)\dotfill 	&	 1.811(16) 	&	 $-$6.07(9) 	&	 $-$2.46(4) 	&	 $-$0.41(9) 	&	 $-$2.69(3) 	&	 9.25(4) 	&	 $-$9.59(8) 	\\
Proper motion in DEC, $\mu_{\delta}$ (mas\,yr$^{-1}$)\dotfill 	&	 $-$10.36(4) 	&	 5.20(10) 	&	 $-$7.33(5) 	&	 $-$9.87(19) 	&	 $-$5.48(6) 	&	 $-$9.58(4) 	&	 $-$8.9(3) 	\\
Parallax, $\pi$ (mas)\dotfill 	&	 0.86(13) 	&	 2.2(11) 	&	 1.1(8) 	&	--	&	 0.5(3) 	&	--	&	 1.84(17) 	\\
Orbital period, $P_b$ (d)\dotfill 	&	 1.198512575218(18) 	&	 4.0835292548(3) 	&	 6.3086296691(5) 	&	 5.2629972182(5) 	&	 12.3271713817(5) 	&	 6.6254930923(13) 	&	 6.83890261536(5) 	\\
Projected semi-major axis, $x$ (lt-s)\dotfill 	&	 1.09144422(6) 	&	 3.0151313(3) 	&	 6.8806577(6) 	&	 3.9828703(4) 	&	 9.2307805(4) 	&	 3.50056678(14) 	&	 10.1641061(3) 	\\
Epoch of ascending node, $T_{\rm ASC}$ (MJD)\dotfill 	&	 50315.26949108(6) 	&	 50273.507005(3) 	&	 50426.28702402(13) 	&	 51396.3661225(3) 	&	47520.4323457(3) 	&	 50442.6431238(4)	&	 50802.29822944(3)	\\
EPS1, $e\sin{\omega_0}$ \dotfill 	&	 3.90(11)$\times 10^{-6}$ 	&	 $-$2.096(17)$\times 10^{-5}$ 	&	 1.60(5)$\times 10^{-6}$ 	&	 2.08(16)$\times 10^{-6}$ 	&	 $-$2.160(5)$\times 10^{-5}$	&	 $-$3.58(8)$\times 10^{-6}$ 	&	$-$6.840(13)$\times 10^{-6}$ 	\\
EPS2, $e\cos{\omega_0}$\dotfill 	&	 3.40(11)$\times 10^{-6}$ 	&	 $-$1.099(16)$\times 10^{-5}$ 	&	 $-$9.20(4)$\times 10^{-6}$ 	&	$-$8.24(16)$\times 10^{-6}$ 	&	 2.46(3)$\times 10^{-6}$	&	 $-$1.165(8)$\times 10^{-5}$ 	&	 $-$1.8059(14)$\times 10^{-5}$ 	\\
First derivative of orbital period, $\dot{P_b}$\dotfill 	&	--	&	--	&	 3.1(15)$\times 10^{-13}$ 	&	--	&	--	&	 7.9(36)$\times 10^{-13}$ 	&	 -- 	\\
First derivative of $x$, $\dot{x}$\dotfill 	&	--	&	--	&	 1.36(16)$\times 10^{-14}$ 	&	--	&	--	&	--	&	8.0(8)$\times 10^{-15}$	\\
Companion mass, $M_c$ ($M_\odot$)\dotfill 	&	--	&	--	&	--	&	--	&	 0.25(3) 	&	--	&	--	\\
Sine of inclination angle, $\sin{i}$\dotfill 	&	--	&	--	&	--	&	--	&	 0.9988(8) 	&	--	&	--	\\
\hline															
\multicolumn{2}{c}{Set Quantities}  			&		&		&		&		&		&		\\
\hline															
Dispersion measure, DM (cm$^{-3}$pc)\dotfill 	&	38.7756	&	58.1438	&	38.0489	&	56.8365	&	13.2984	&	31.8509	&	8.99761	\\
\hline															
\multicolumn{2}{c}{Derived Quantities} 			&		&		&		&		&		&		\\
\hline															
$\log_{10}$(Characteristic age, yr) \dotfill 	&	9.7	&	9.83	&	10.18	&	9.77	&	9.68	&	9.45	&	9.93	\\
$\log_{10}$(Surface magnetic field strength, G) \dotfill 	&	8.24	&	8.57	&	8.69	&	8.44	&	8.5	&	8.45	&	8.84	\\
$\log_{10}$(Edot, ergs/s) \dotfill 	&	34.12	&	33.22	&	32.28	&	33.57	&	33.66	&	34.2	&	32.45	\\
Epoch of periastron, $T_0$ (MJD)\dotfill 	&	 50315.432(4) 	&	 50276.256(5) 	&	 50429.268(5) 	&	 51398.790(16) 	&	 47529.900(3) 	&	 50446.270(7) 	&	 50806.1118(8) 	\\
Orbital eccentricity, $e$\dotfill 	&	 5.18(11)$\times 10^{-6}$ 	&	 2.367(17)$\times 10^{-5}$ 	&	 9.35(5)$\times 10^{-6}$ 	&	 8.50(16)$\times 10^{-6}$ 	&	 2.174(5)$\times 10^{-5}$ 	&	 1.219(8)$\times 10^{-5}$ 	&	 1.9311(14)$\times 10^{-5}$ 	\\
Longitude of periastron, $\omega_0$ (deg)\dotfill 	&	 48.9(12) 	&	 242.3(4) 	&	 170.1(3) 	&	 165.8(11) 	&	 276.49(7) 	&	 197.1(4) 	&	 200.75(4) 	\\
Pulsar mass, $M_P$ ($M_\odot$)\dotfill	&	--	&	-- & -- & -- & 1.5(0.2) & -- & -- \\
$i$ limit from $\dot{x}$ measurement (degrees)\dotfill & -- & -- & $\leq 31$ & -- & -- & -- & $\leq 69$\\
\hline																											
\end{tabular}
\end{adjustwidth}
\end{table*}
\end{landscape}

%
\clearpage 
\pagestyle{empty}
\oddsidemargin=1.4cm
\evensidemargin=1.4cm
\topmargin=5.2cm
\begin{landscape}
\centering
\begin{table*}
\centering
\small
\caption{Parameters for the binary pulsars described by DD binary model, PSRs  J1022+1001, J1600$-$3053, and J1643$-$1224. Numbers in brackets are the \textsc{tempo2} 1-sigma uncertainties on the last quoted decimal place, including split-Cholesky analysis.}
\begin{tabular}{llll}
\hline					
\hline					
Pulsar name\dotfill 	&	 J1022+1001  	&	 J1600$-$3053  &	 J1643$-$1224	\\
MJD range\dotfill 	&	 52649.7---55618.6  	&	 52302.0---55618.8  &	 49421.8---55618.9	\\
Data span (yr)\dotfill 	&	8.13	&	9.08	&	16.97\\
Number of TOAs\dotfill 	&	615	&	715	&	488\\
Rms timing residual ($\mu s$)\dotfill 	&	1.8	&	0.8	&	2.8\\
\hline					
\multicolumn{2}{c}{Measured Quantities}  		&		&\\
\hline					
Right ascension (RA), $\alpha$ (hh:mm:ss)\dotfill 	&	  10:22:58.0007(13) 	&	  16:00:51.903452(7) 	&	  16:43:38.160985(9) \\
Declination (DEC), $\delta$ (dd:mm:ss)\dotfill 	&	 +10:01:52.77(5) 	&	 $-$30:53:49.3653(3) &	 $-$12:24:58.6783(6) 	\\
Pulse frequency, $\nu$ (s$^{-1}$)\dotfill 	&	 60.7794479636137(3) 	&	 277.9377070213120(12) 	&	 216.373337179973(9) 	\\
First derivative of pulse frequency, $\dot{\nu}$ (s$^{-2}$)\dotfill 	&	 $-$1.60095(6)$\times 10^{-16}$ 	&	 $-$7.3385(4)$\times 10^{-16}$ 	&	 $-$8.6433(5)$\times 10^{-16}$ \\
Proper motion in RA, $\mu_{\alpha} \cos \delta$ (mas\,yr$^{-1}$)\dotfill 	&	 $-$17.09(3) 	&	 $-$0.99(4) &	 5.94(5)	\\
Proper motion in DEC, $\mu_{\delta}$ (mas\,yr$^{-1}$)\dotfill 	&	--	&	 $-$7.22(15) 		&	 3.94(18) \\
Parallax, $\pi$ (mas)\dotfill 	&	 1.1(3) 	&	 0.48(11) &	 1.27(19) 	\\
Orbital period, $P_b$ (d)\dotfill 	&	 7.8051360(16) 	&	 14.3484577721(3) 	&	 147.01728(7) \\
Projected semi-major axis, $x$ (lt-s)\dotfill 	&	 16.765395(14) 	&	 8.8016536(13) 	&	 25.0726150(7) \\
Epoch of periastron, $T_0$ (MJD)\dotfill 	&	 49778.4080(11) 	&	 53295.5390(7) &	 49577.972(3) 	\\
Orbital eccentricity, $e$\dotfill 	&	 9.683(17)$\times 10^{-5}$ 	&	 1.73729(10)$\times 10^{-4}$ &	 5.05753(9)$\times 10^{-4}$ 	\\
Longitude of periastron, $\omega_0$ (deg)\dotfill 	&	 97.64(5) 	&	 181.832(17) 	&	 321.857(6) \\
First derivative of orbital period, $\dot{P_b}$\dotfill 	&	 5.5(23)$\times 10^{-13}$ 	&	 -- & --	\\
First derivative of $x$, $\dot{x}$ \dotfill 	&	 1.15(16)$\times 10^{-14}$ 	&	$-$4.2(7)$\times 10^{-15}$ & $-$5.25(16)$\times 10^{-14}$	\\
Periastron advance, $\dot{\omega}$ (deg/yr)\dotfill 	&	 0.012(4) 	&	--	& $-$0.0007(4)\\
Companion mass, $M_c$ ($M_\odot$)\dotfill 	&	 2.2(2.4) 	&	 0.34(15) & --	\\
Sine of inclination angle, $\sin{i}$\dotfill 	&	 0.69(18) 	&	 0.87(6)  & --	\\
\hline					
\multicolumn{2}{c}{Set Quantities}  		&		&\\
\hline					
Dispersion measure, DM (cm$^{-3}$pc)\dotfill 	&	10.2531	&	52.3249	&	62.4143	\\
\hline					
\multicolumn{2}{c}{Derived Quantities} 		&		&\\
\hline					
$\log_{10}$(Characteristic age, yr) \dotfill 	&	9.78	&	9.78	&	9.6\\
$\log_{10}$(Surface magnetic field strength, G) \dotfill 	&	8.93	&	8.27	&	8.47\\
$\log_{10}$(Edot, ergs/s) \dotfill 	&	32.58	&	33.91	&	33.87\\
Pulsar mass, $M_P$ ($M_\odot$)\dotfill	&	--	&	2.4(1.7)	& --	\\
$i$ limit from $\dot{x}$ measurement (degrees)\dotfill & $\leq 84$ & $\leq 67$ & $\leq 28$\\
\hline																														
\end{tabular}
\end{table*}
\end{landscape}
\clearpage 
\pagestyle{empty}
\oddsidemargin=1.4cm
\evensidemargin=1.4cm
\topmargin=5.2cm
\begin{landscape}
\centering
\begin{table*}
\centering
\small
\caption{Parameters for the binary pulsars described by T2 binary model, PSRs  J0437$-$4715, J1713+0747, and J1909$-$3744. Numbers in brackets are the \textsc{tempo2} 1-sigma uncertainties on the last quoted decimal place, including split-Cholesky analysis. In each case where a companion mass is measured from the Shapiro delay, the corresponding $\sin{i}$ parameter is linked to the Kopeikin parameter, $i$.}
\begin{tabular}{llll}
\hline									
\hline									
Pulsar name\dotfill 	&	 J0437$-$4715  	  	&	 J1713+0747  	&	 J1909$-$3744  	\\
MJD range\dotfill 	&	 50191.0---55619.2  	  	&	 49421.9---55618.9  	&	 52618.4---55619.1  	\\
Data span (yr)\dotfill 	&	14.86		&	16.97	&	8.22	\\
Number of TOAs\dotfill 	&	5065		&	622	&	1368	\\
Rms timing residual ($\mu s$)\dotfill 	&	0.3		&	0.4	&	0.2	\\
\hline									
\multicolumn{2}{c}{Measured Quantities}  			&		&				\\
\hline									
Right ascension (RA), $\alpha$ (hh:mm:ss)\dotfill 	&	  04:37:15.8961737(6) 		&	  17:13:49.5327220(19) 	&	  19:09:47.4346749(11) 	\\
Declination (DEC), $\delta$ (dd:mm:ss)\dotfill 	&	 $-$47:15:09.110714(7) 		&	 +07:47:37.49795(6) 	&	 $-$37:44:14.46674(5) 	\\
Pulse frequency, $\nu$ (s$^{-1}$)\dotfill 	&	 173.6879458121843(5) 	&	 218.8118404348011(11) 	&	 339.3156872882446(3) 	\\
First derivative of pulse frequency, $\dot{\nu}$ (s$^{-2}$)\dotfill 	&	 $-$1.728361(5)$\times 10^{-15}$ 		&	 $-$4.08380(6)$\times 10^{-16}$ 	&	 $-$1.614817(5)$\times 10^{-15}$ 	\\
Proper motion in RA, $\mu_{\alpha} \cos \delta$ (mas\,yr$^{-1}$)\dotfill 	&	 121.4385(20) 	 	&	 4.912(7) 	&	 $-$9.517(5) 	\\
Proper motion in DEC, $\mu_{\delta}$ (mas\,yr$^{-1}$)\dotfill 	&	 $-$71.4754(20) 	&	 $-$3.888(14) 	&	 $-$35.797(17) 	\\
Parallax, $\pi$ (mas)\dotfill 	&	 6.37(9) 		&	 0.86(9) 	&	 0.810(3) 	\\
Orbital period, $P_b$ (d)\dotfill 	&	 5.7410459(4) 		&	 67.825130978(4) 	&	 1.533449474406(13) 	\\
Projected semi-major axis, $x$ (lt-s)\dotfill 	&	 3.36671444(5) 		&	 32.3424210(5) 	&	 1.89799118(4) 	\\
Epoch of periastron, $T_0$ (MJD)\dotfill 	&	 54501.4671(3) 		&	 51997.5804(9) 	&	 53631.39(4) 	\\
Orbital eccentricity, $e$\dotfill 	&	 1.91811(15)$\times 10^{-5}$ 		&	 7.49373(17)$\times 10^{-5}$ 	&	 1.14(10)$\times 10^{-7}$ 	\\
Longitude of periastron, $\omega_0$ (deg)\dotfill 	&	 1.363(17) 		&	 176.201(5) 	&	 156(8) 	\\
First derivative of orbital period, $\dot{P_b}$\dotfill 	&	 3.728(6)$\times 10^{-12}$ 	&	 1.7(7)$\times 10^{-12}$ 	&	 5.03(6)$\times 10^{-13}$ 	\\
Periastron advance, $\dot{\omega}$ (deg/yr)\dotfill 	&	 0.0138(13) 	&	--	&	--	\\
Companion mass, $M_c$ ($M_\odot$)\dotfill 	&	 0.224(7) 		&	 0.34(5) 	&	 0.2067(19) 	\\
Inclination angle, $i$ (degrees)\dotfill 	&	 137.56(4) 		&	 69(3) 	&	 93.52(9) 	\\
Longitude of ascending node, $\Omega$ (degrees)\dotfill 	&	 207.0(12) 		&	 99(4) 	&	 39(10) 	\\
\hline									
\multicolumn{2}{c}{Set Quantities}  			&		&				\\
\hline									
Dispersion measure, DM (cm$^{-3}$pc)\dotfill 	&	2.64498	&	15.9903	&	10.3932	\\
\hline									
\multicolumn{2}{c}{Derived Quantities} 			&		&				\\
\hline									
$\log_{10}$(Characteristic age, yr) \dotfill 	&	9.2		&	9.93	&	9.52	\\
$\log_{10}$(Surface magnetic field strength, G) \dotfill 	&	8.76		&	8.3	&	8.31	\\
$\log_{10}$(Edot, ergs/s) \dotfill 	&	34.07		&	33.55	&	34.34	\\
Pulsar mass, $M_P$ ($M_\odot$)\dotfill	&	1.44(7)	&	1.7(4)	&	1.47(3)	\\
\hline																		
\end{tabular}
\end{table*}
\end{landscape}
\end{document}